\documentclass[12pt,a4paper]{article}

\title{\textbf{A natural electroweak scale}}
\author{Lucila Z\'arate}
\date{}

\usepackage{amsmath, amssymb, amsfonts, mathrsfs, graphicx, fancyhdr}
\usepackage[square,sort,comma,numbers]{natbib}
\usepackage[pdftex]{color}
\def\be{\begin{equation}}
\def\ee{\end{equation}}
\def\ba{\begin{align}}
\def\ea{\end{align}}

\begin{document}

\begin{titlepage}

\vspace*{0.0cm}
\begin{flushright}
ZMP-HH/15-12
\end{flushright}

\vskip 1cm 

\begin{center}
{\bf Hiding the little hierarchy problem in the NMSSM}

\vspace{1cm}

\textbf{
Jan Louis$^{a,b}$ and
Lucila Zarate$^{a}$
}
\\[5mm]

$^{a}${\em Fachbereich Physik der Universit\"at Hamburg, Luruper Chaussee 149, 22761 Hamburg, Germany}
\vskip 0.3cm
\vskip 0.3cm

{}$^{b}${\em Zentrum f\"ur Mathematische Physik,
Universit\"at Hamburg,\\
Bundesstrasse 55, D-20146 Hamburg, Germany}
\vskip 0.3cm

\end{center}

\vspace{1cm}

\begin{abstract}

\noindent
In this paper we consider a set of soft supersymmetry
breaking terms within the
NMSSM which leads to a small hierarchy between the supersymmetry
breaking scale and the electroweak scale. Specifically
only  the gaugino masses and
the soft term in the Higgs sector are non-vanishing at the GUT scale.
This pattern can be found in gaugino mediated models and in higher-dimensional
orbifold GUTs.
We study the phenomenology of this scenario and find different low energy
spectra  depending on the Yukawa coupling $\lambda$ of the NMSSM
singlet. 
In particular, for low values of $\lambda$ the singlet is the lightest
scalar and the singlino is the LSP while for large values of $\lambda$
both are heavy and the gravitino can be the LSP. 
The singlet pseudoscalar is very light in a broad range of the parameter space.
%{{\bf In addition, we compute the soft terms for global supersymmetric theories in a model independent way and apply these to an example that motivates the set of soft terms studied in this work.}}

\end{abstract}

\vfill

June 2015

\end{titlepage}

\section{Introduction}

The discovery of a $126\,$GeV Higgs boson at the LHC
\cite{Aad:2012tfa,Chatrchyan:2012ufa} together with the absence of a
signal for supersymmetric particles up to the $\text{TeV}$ scale
\cite{Chatrchyan:2013lya,Aad:2013wta}
generated some tension in supersymmetric theories. The mass of the
Higgs boson is consistent with the range predicted by the minimal
supersymmetric standard model (MSSM) but the supersymmetry breaking
scale ${M}_{\rm s}$ is pushed far above the weak scale $M_z$
leading to the {\it little hierarchy problem}. For recent reviews on the
status of supersymmetry after the LHC see, for example, \cite{Feng:2013pwa, Craig:2013cxa}. 

In the MSSM the tree level Standard Model Higgs mass is bounded from
above by the Z boson mass $(M_z\simeq91\text{GeV})$ which is reached 
for  large values of $\tan\beta$, defined as the ratio of the two Higgs 
vacuum expectation values (VEV) and taken as a free parameter in the MSSM.
Therefore radiative corrections need to be large in order to achieve
the measured $126\,\text{GeV}$ value which in turn put a lower bound on
the soft  supersymmetry breaking parameters of order ${\cal
  O}(\text{TeV})$. In principle this fact could be used to argue
for the absence of signals of supersymmetry at the LHC. However, soft
parameters trigger electroweak symmetry breaking through the following
 tree level  relation \cite{Martin:1997ns}
\be M_z^2=2\ (-\mu^2+\hat{m}^2)\label{Mz_mssm}\ ,\ee
where $\mu$ is a free parameter in the MSSM  and $\hat{m}$ is fixed by
the soft parameters and $\tan\beta$. From \eqref{Mz_mssm} one learns
that  a cancellation between $\mu$ and $\hat{m}$ is needed in
order to match $M_z$ with its experimental value. This fact
is the above mentioned little hierarchy problem. 
Various suggestions towards its solutions can be found
 for example in \cite{Horton:2009ed,Feng:1999zg,Brummer:2013dya,Batra:2003nj}. 
%and \cite{ArkaniHamed:2004fb,Giudice:2004tc}.
%Different ideas were proposed to solve the problem, among them,  non universal gaugino masses \cite{Horton:2009ed},focus points \cite{Feng:1999zg}, special relations between gaugino and soft scalar masses,\cite{Brummer:2013dya}, extended gauge groups\cite{Batra:2003nj} and split susy\cite{ArkaniHamed:2004fb},\cite{Giudice:2004tc}.

It is of interest to address the problem within non-minimal versions of the supersymmetric standard model. In particular, in the Next-to-Minimal
Supersymmetric Standard Model (NMSSM)  a singlet chiral multiplet is
added to the field content of the MSSM (for a comprehensive review see
\cite{Ellwanger:2009dp}). The singlet couples to the Higgs sector of
the MSSM via a Yukawa interaction $\lambda$. The
latter also generates an effective $\mu$ term that elegantly solves the
$\mu$-problem for the scale  (or  $\mathbb{Z}_3$) invariant
  version of the NMSSM.\footnote{{In the more general case 
  additional sources for the $\mu$ term exist and thus spoil this property.
  However, there are well motivated generalized versions of the scale
  invariant NMSSM
 for which the $\mu_h$ term can be generated after supersymmetry breaking and can be naturally of order of the soft parameters \cite{Lee:2011dya}.}}
Furthermore,  $\lambda$ provides an additional
quartic term for the Higgses which enhances the tree level Higgs mass
with respect to the MSSM bound for large values of $\lambda$ and low  values of
$\tan\beta$. In this regime, the relevant parameters  that  determine
$\hat{m}$ differ from the MSSM case and thus 
can lead to new possibilities to study the little hierarchy.
%{\color{red} In this regime, the relevant parameters  that  determine $\hat{m}$ differ from the MSSM case and thus 
%can relieve the fine tuning problem \cite%{King:2012tr,Kang:2012sy,King:2012is,Agashe:2012zq}. }
%For examples with the $\lambda$ coupling above the perturbative range see \cite{Gherghetta:2012gb}. {\bf L: mention it solves the mu problem}

In the present paper we investigate the question whether a special
configuration of 
the soft parameters can explain the little hierarchy problem  within
the NMSSM. In particular we consider scenarios where  $\hat{m}$ is suppressed with respect to the supersymmetry breaking scale, i.e.
\be \hat{m}\ll {M}_{\mathrm{s}}\label{mnolhp}\ .
\ee 
 $\hat{m}$ depends on the soft terms defined at
 the GUT scale. 
%These encode the knowledge of the UV interactions and, from a bottom up approach, they are constraint by the requirements on the low energy theory. 
%Some of the prominent UV-completions are string theories or, as an intermediate step, supersymmetric field theories in a higher-dimensional space-time. 
For the customary choice of universal boundary conditions at the GUT scale, $\hat{m}$ is determined by the gaugino masses $M_0$ and soft scalar masses $m_0$.\footnote{Universal boundary conditions at the GUT scale correspond to the minimal setup of soft terms $M_0,m_0,A,b$. $\hat{m}$ depends mildly on the $A$-terms thus they can be ignored in the argument.}
If $M_0$ and $m_0$ are related to yield \eqref{mnolhp} 
then the little hierarchy problem disappears.\footnote{Note, however, that this does not imply that the fine tuning is relieved. In order to have this suppression one requires a very precise relation among the soft parameters and small deviations from this value would spoil the necessary cancellations.
Therefore, such a relation should be understood as an outcome of  a UV completed theory.}
This scenario was studied in the MSSM in \cite{Brummer:2013dya}, see also \cite{Harigaya:2015iva,Harigaya:2015jba}. 
Here the relation is derived in singlet extensions of the MSSM with non-universal Higgs masses at the GUT scale. 
The latter is a natural option to generalize the minimal setup,  the
Higgses carry no family index and thus there is no danger of flavor changing neutral currents (FCNC) and no urge for the universality condition. 

In particular, we consider soft gaugino masses and soft terms in
the Higgs
sector while all other soft terms vanish.
This setup can be embedded in e.g.~higher-dimen\-sional orbifold GUTs  where the singlet together with 
the quarks and leptons 
 are confined to a four-dimensional subspace (brane or orbifold fixed
 point) 
while the  gauge fields and the Higgses propagate in the bulk.\footnote{See \cite{Kawamura:2000ev,
  Altarelli:2001qj, Hebecker:2001wq, Asaka:2001eh, Hall:2001xr} for examples of higher dimensional orbifold GUTs in five
and six dimensions and
\cite{Kobayashi:2004ud,Forste:2004ie,Kobayashi:2004ya, Buchmuller:2007qf} for examples 
derived from asymmetric orbifold compactifications in the heterotic
string theory.}
The additional assumption that supersymmetry breaking occurs at a spatially separated brane by a hidden field \cite{Randall:1998uk} leads to the so called gaugino mediation \cite{Chacko:1999mi, Chacko:1999hg, Schmaltz:2000ei} and reproduces the boundary conditions for the soft terms studied in this paper. 

In the present work we study the phenomenology of these scenarios. The low energy spectrum can be very different from the MSSM and has implications for the next LHC run. The predictions depend on the value of the Yukawa coupling $\lambda$. 
For $\lambda\lesssim\mathcal{O}(10^{-1})$ the singlet scalar and singlino are  heavy $\mathcal{O}(\text{TeV})$ while for lower values $\lambda\lesssim \mathcal{O}(10^{-4})$ the singlet becomes the lightest scalar and the singlino the LSP. 
The pseudoscalar singlet is $\mathcal{O}(100) \text{GeV}$ below $\lambda\lesssim\mathcal{O}(10^{-1})$ and reaches $\mathcal{O}(\text{TeV})$ for larger values of $\lambda$.
%The pseudoscalar singlet is below $\mathcal{O}(100) \text{GeV}$ in all of the parameter space. 
The gravitino mass is $\mathcal{O}(10)\text{GeV}$ and can be the LSP depending on the value of $\lambda$. These scenarios can also be interesting for dark matter searches. 

In addition we study whether the special relation among the soft terms computed above could be obtained from a more fundamental theory. 
Soft terms at the GUT scale rely on the mechanism and mediation of supersymmetry breaking.
%, hence their pattern can be very restrictive. 
Within supergravity, i.e.~for gravity mediated supersymmetry breaking, soft terms are computed in \cite{Kaplunovsky:1993rd} in a model independent way. 
More precisely, without specifying the dynamics that trigger
supersymmetry breaking, they are parametrized through general (unknown) couplings in the K\"{a}hler potential, the superpotential and the gauge kinetic functions. 
One can also consider, as an intermediate step of a UV completion,
effective globally supersymmetric theories. Supersymmetry breaking can
be communicated for example, via gauge or gaugino
mediation. In this paper, pursuing the spirit of
\cite{Kaplunovsky:1993rd} we compute the soft terms for this class of
theories in a model independent way.  
We then use the example suggested in \cite{Brummer:2013dya} to provide
the special relation between soft gaugino and Higgs scalar masses
required to ease the little hierarchy problem. 
It is worth stressing that this relation should come out of the UV theory, adjusting the coefficient to the necessary value implies a fine tuning as severe as in conventional MSSM models. However, the aim of the example is to illustrate that one can expect the little hierarchy problem to be an artifact of our ignorance and to hint for patterns of soft terms at the GUT scale that lead to a natural electroweak scale.

%{ However, the fine tuning measure rely on the assumption that soft parameters are independent variables, our approach is instead to assume soft terms are related to yield a little hierarchy, and calculate this relation. This leads to new perspectives to find natural models.
%We propose scenarios that would be naively considered exluded  by the requirements of absence of fine tuning 
%but can develop a little hierarchy. 
%The first rely on the assumption that soft parameters are independent variables, our approach is instead to assume soft terms are related to yield a little hierarchy, and calculate this relation. This opens a new perspective to find natural models.
%The little hierarchy problem is not solved until one finds a UV completion capable of realising it. 

%However, the aim of the example is to illustrate that one could expect the little hierarchy to be an artifact of our ignorance and a hint for patterns of soft terms at the GUT scale that lead to a natural electroweak scale.

This paper is organized as follows. In section \ref{EWSB_nmssm} we introduce the relevant parameters in the NMSSM, in section \ref{calculationk} we compute the relation between the soft gaugino and Higgs scalar masses that leads to the condition \eqref{mnolhp} and in section \ref{pheno} we investigate the phenomenological implications of the models studied. In section \ref{HDOG} we motivate via higher-dimensional orbifold GUTs the structure of soft terms used in section \ref{gauginoscalarrelation} and compute them within a simple example that yields the required values. In Appendix \ref{appendixsoftterms} we provide the soft terms for effective global supersymmetric theories, used in section~\ref{HDOG}.

\section{A low electroweak scale from a special gaugino-scalar mass relation}\label{gauginoscalarrelation}

\subsection{Conditions for electroweak symmetry breaking}\label{EWSB_nmssm}

In this section we present the Lagrangian of the NMSSM and the conditions for
electroweak symmetry breaking. The NMSSM extends the MSSM by adding
a chiral gauge singlet supermultiplet $S$. The singlet is coupled to
the Higgs sector via  a Yukawa interaction with coupling $\lambda$ and
it contributes to the electroweak symmetry breaking as an additional
Higgs. To be more precise, the singlet gets a vacuum expectation value
(VEV) and mixes with the MSSM Higgses in the mass matrix. 
The NMSSM superpotential is given by
\begin{equation}\begin{aligned} W_{\mathrm{NMSSM}}=&(\lambda S +\mu_h)H_u H_d+\tfrac{1}{3}\kappa S^3+\\
&\sum_{\text{generations}}y_u Q
U_RH_u+y_dQ_LD_RH_d+y_eL_LE_RH_d\ ,\label{Wnmssm}
\end{aligned}\end{equation}
where $S$ is the NMSSM singlet and $H_u,H_d$ are the MSSM Higgs
multiplets. $\lambda$ and $\kappa$ are dimensionless Yukawa couplings
and $\mu_h$ is the supersymmetric Higgs mass term.\footnote{{In the literature the
  NMSSM often denotes  the scale (or $\mathbb{Z}_3$) invariant version
  of singlet extensions of the MSSM which has $\mu_h=0$. 
 %Here we study both possibilities, $\mu_h=0$ and $\mu_h\neq 0$. 
 % The possible \textbf{}origin of the $\mu_h$ term is discussed in Section~\ref{HDOG}.
In the generalized versions considered in \cite{Lee:2011dya} an
 underlying symmetry forbids the $\mu_h$ term  before supersymmetry
 breaking. However after supersymmetry breaking this can be
 non-vanishing and naturally be of order of the soft parameters. One could also consider quadratic and linear terms for the singlet to be generated after superymmetry breaking. In this work we take these to be vanishing, this choice is motivated in section \ref{HDOG}.
 }}
$y_u,y_d,y_e$ are  the Yukawa couplings of the MSSM, $Q$ are the quark doublets, $U_R$ and $D_R$ are the quark singlets, $L_L$ are the lepton doublets and $E_R$ are the lepton singlets.
In addition to the supersymmetric interactions given in
\eqref{Wnmssm}, supersymmetry breaking induces soft terms
given by
\begin{equation}\begin{aligned}
V_{\text{soft}}=&\tfrac{1}{2}M_a\lambda_a\bar{\lambda}_a+m_{h_u}^2\vert H_u\vert^2+m_{h_d}^2\vert H_d\vert^2+m_{s}^2\vert S\vert^2\\
&+(\lambda A_{\lambda}H_u  H_dS+\tfrac{1}{3}\kappa A_{\kappa}S^3+\frac{1}{2}b_sS^2+\xi_sS+b_h\ H_u  H_d+\text{h.c.})\\
&+\sum_{\text{generations}}m_{q}^2\vert Q\vert^2+m_{u}^2\vert U_R\vert^2+m_{d}^2\vert D_R\vert^2+m_{l}^2\vert L\vert^2+m_{e}^2\vert E\vert^2\\
&+(y_{u} A_{u} Q  H_u U_R- y_{d} A_{d} Q  H_d D_R-y_{e}
{A_{e}} L  H_dE_R+\text{h.c.}) \ ,
\end{aligned}\end{equation}
%\begin{equation}\begin{aligned}
%V_{\text{soft}}=&\tfrac{1}{2}M_a\lambda_a\bar{\lambda}_a+m_{h_u}^2\vert H_u\vert^2+m_{h_d}^2\vert H_d\vert^2+m_{s}^2\vert S\vert^2\\
%&+(\lambda A_{\lambda}H_u  H_dS+\tfrac{1}{3}\kappa A_{\kappa}S^3+b_h\ H_u  H_d+\text{h.c.})\\
%&+\sum_{\text{generations}}m_{q}^2\vert Q\vert^2+m_{u}^2\vert U_R\vert^2+m_{d}^2\vert D_R\vert^2+m_{l}^2\vert L\vert^2+m_{e}^2\vert E\vert^2\\
%&+(y_{u} A_{u} Q  H_u U_R- y_{d} A_{d} Q  H_d D_R-y_{e}
%{A_{e}} L  H_dE_R+\text{h.c.}) \ ,
%\end{aligned}\end{equation}
where $M_a,a=1,2,3 $ are the three soft gaugino masses, $m_j$
are the soft scalar masses, $A_{y}$ are the $A$-terms, $b_h$ is the 
$b$-term and $b_s,\xi_s$ are b-term and tadpole soft terms of the singlet.\footnote{
{Provided that $\mu_h$ and $b_h$ are non-vanishing after supersymmetry breaking the quadratic and linear soft terms of the singlet
%The $\mu_h$ term slightly breaks the $\mathbb{Z}_3$ symmetry of the potential. 
 can grow radiatively and thus must be included in the discussion of the potential.}}

After these preliminaries the computation of the scalar potential is
straightforward. For the Higgs sector together with the singlet
the potential reads~\cite{Ellwanger:2009dp}
\begin{equation}\begin{aligned}
V_{\text{higgs}}=&\frac{1}{8}(g_1^2+g_2^2)(\vert h_u\vert^2-\vert h_d\vert^2)^2+
(m_{h_u}^2+\mu^2)\vert h_u\vert^2+(m_{h_d}^2+\mu^2)\vert h_d\vert^2\\
&+\lambda^2 \vert h_u\vert^2 \vert h_d\vert^2+\kappa^2 \vert s\vert^4+m_s^2\vert s\vert^2+(-b\,h_u h_d+\frac{\kappa}{3}
A_{\kappa}s^3+\frac{1}{2}b_ss^2+\xi_ss+\text{h.c}),
\label{V_higgs}\end{aligned}\end{equation}
%\begin{equation}\begin{aligned}
%V_{\text{higgs}}=\frac{1}{8}(g_1^2+&g_2^2)(\vert h_u\vert^2-\vert h_d\vert^2)^2+
%(m_{h_u}^2+\mu^2)\vert h_u\vert^2+(m_{h_d}^2+\mu^2)\vert h_d\vert^2\\
%-2b&\ (h_u h_d+\text{h.c.}) +\lambda^2 \vert h_u\vert^2 \vert h_d\vert^2
%+\kappa^2 \vert s\vert^4+m_s^2\vert s\vert^2+\frac{\kappa}{3}
%A_{\kappa}(s^3+\text{h.c})\ ,
%\label{V_higgs}\end{aligned}\end{equation}
where $h_u, h_d,s$ are the scalar components of the respective
supermultiplets
and $\mu$ and $b$ are defined as
\begin{equation}\begin{aligned}
\mu&=\mu_{\text{eff}}+\mu_h,\quad \ \mu_{\text{eff}}=\lambda s\label{defmu}\,,\qquad
b=\mu_{\text{eff}}\, b_{\text{eff}} +b_h^2,\quad
b_{\text{eff}}=A_{\lambda}+\kappa s\,.
%\label{b_nmssm}
\end{aligned}\end{equation}

The VEVs  $\langle h_u\rangle,\langle h_d\rangle, \langle s \rangle$
 can be computed by minimizing \eqref{V_higgs}. Since the soft terms 
are of order of the supersymmetry breaking scale
$M_{\mathrm{s}}$ and we assume $M_{s}\gg\langle h_u\rangle, \langle
h_d\rangle$ the VEV of the singlet can be obtained from the minimum of $V_s$ given by
\be V_s\simeq\kappa^2s^4+\frac{2}{3}\kappa
A_{\kappa}s^3+(m_s^2+b_s)s^2+2\xi_ss \label{Vs}\ .\ee
%by solving \be 2\kappa^2s^3+\kappa A_{\kappa}s^2+(m_s^2+b_s)s+\xi_s=0\ .\ee
The global minimum, for $(m_s^2+b_s)<0$ and neglecting $A_{\kappa}$, which corresponds to our parameter space as explained in section \ref{calculationk},
can be approximated by
\be \langle s\rangle \simeq -2
\text{sign}(q)\sqrt{\frac{-p}{3}}\cosh\Big(\frac{1}{3}\text{arccosh}(\sqrt{-x})\Big)\ ,\label{svev}\ee
with $p=(m_s^2+b_s)/(2\kappa^2)$, $q=\xi_s/(2\kappa^2)$ and $x=27q^2/(4p^3)$. 
%\be \langle s\rangle \simeq -
%\frac{A_{\kappa}}{4\kappa}-\frac{1}{4\kappa}\sqrt{A_{\kappa}^2-8m_s^2}\
%.
%\ee
The remaining two minimization conditions  determine $M_z$ and $\tan\beta:=\frac{<h_u>}{<h_d>}$, via the following equations
\be M_z^2=2\ (-\mu^2+\hat m^2)\label{Mz}\,,\ee
where we defined
\be \hat m^2:= \frac{m_{h_d}^2-\tan^2\beta m_{h_u}^2}{\tan^2\beta-1}\,,\label{m}\ee 
and
\be \sin(2\beta)=\frac{2b}{m_{h_u}^2+m_{h_d}^2+2\mu^2+\lambda^2
  v^2}\quad \label{sin2b}\ ,\ee
with $v^2=\langle h_u\rangle^2+\langle h_d\rangle^2=(174\text{GeV})^2,\,
\beta\in[\frac{\pi}{4},\frac{\pi}{2}]$ and all parameters are to be
taken at 
the scale $M_{\mathrm{s}}$. 
%As anticipated in the introduction we look for the special relation between gaugino and  scalar masses that induces a low Fermi scale. More explicitely, we define 
%\be M=km_0\ee
%and calculate the $k=k(\tan\beta)$ for which $m$, in \eqref{m}, satisfies
%\be m\ll M_{\text{s}}.\ee
 The value of $\tan\beta$ and the (running) top mass $m_t$ fix the top Yukawa coupling via
%The value of $\tan\beta$ fixes the top Yukawa coupling via the measured (running) top mass given as follows
 \be m_t=y_t h_u=y_t v sin\beta\,.\label{mtop}\ee

The scalar potential gets threshold corrections at one loop which
can be computed from the Coleman-Weinberg potential \cite{Ellwanger:2009dp}
\be \Delta V=\frac{1}{64\pi^2}\,\text{Str}\
M^4\left(\log\left(\frac{M^2}{M_s^2}\right)-\frac{3}{2}\right),\ee 
where $M_{\mathrm{s}}=\sqrt{m_{\tilde{t}_1}m_{\tilde{t}_2}}$ with
$m_{\tilde{t}_{1,2}}$ being the eigenvalues of the stop mass matrix. They explicitely read
\be m_{\tilde{t}_{1,2}}^2=m_{t}^2+\tfrac{1}{2}(m_{q}^2+m_{u}^2)\mp
\sqrt{W}\label{stops}\ ,\ee
where 
%$m_t$ is the top mass and 
$W=\frac{1}{4}(m_{q}^2-m_{u}^2)^2+m_t^2(A_u-\mu\ h_d/h_u)^2$ is the mixing parameter. 
The dominant contribution in $\Delta V$ comes from the top sector 
% L: since it has the largest coupling to the Higgs field ($h_u$) 
and shifts the soft Higgs masses as follows 
\be m_{h_u}^2\to m_{h_u}^2+\frac{3}{32\pi^2}y_t^2 c_{\tilde{t}}\
,\quad m_{h_d}^2\to m_{h_d}^2\ ,\quad m_s^2\to m_s^2\ ,
\label{thresholdsofthiggs}\ee
where
 \be\qquad c_{\tilde{t}}=m_{\tilde{t}_1}^2\left( \log
    \frac{m_{\tilde{t}_1}^2}{M_s^2}-1\right)+m_{\tilde{t}_2}^2\left(\log
    \frac{m_{\tilde{t}_2}^2}{M_{\mathrm{s}}^2}-1\right)-2m_{t}^2\left(\log
    \frac{m_{t}^2}{M_{\mathrm{s}}^2}-1\right)\ .\ee  

 Before continuing let us recall a few properties of the Higgs mass
 within the NMSSM. The three neutral scalars $h_u,h_d,s$ mix in a $3\times3$ mass
 matrix that should be diagonalized in order
 to obtain the three mass eigenstates~\cite{Ellwanger:2009dp}. 
For most of the parameter space
%In the MSSM, the lightest Higgs is typically $h_u$-like and the heavier Higgs is sufficiently large to avoid any mixing. In the NMSSM this picture could change, one possibilty is that the lightest particle is singlet-like and yet allows for a SM-like higgs mass with its observed value of $126\text{GeV}$. 
%(However we do not encounter this scenario through this work.) 
the singlet is heavy and a SM-like Higgs is found when the mixing
%between all three scalars $h_u,h_d,s$ 
with with the singlet can be neglected. In this case 
one finds for the mass of the  SM-like Higgs
at one loop  \cite{Ellwanger:1993hn}
%\be m_h^2\simeqM_z^2\cos^22\beta+\lambda^2v^2\sin^22\beta+\frac{3m_t^4}{4\pi v^2}\left(\ln\frac{M_{\mathrm{s}}^2}{m_t^2}+\frac{X_t^2}{M_{\mathrm{s}}^2}\left(1-\frac{X_t^2}{12M_{\mathrm{s}}^2}\right)\right)\label{mh}\ee
\be 
m_h^2\simeq M_z^2\cos^22\beta+\lambda^2v^2\sin^22\beta+
\frac{3m_t^4}{4\pi^2 v^2}\ln\frac{M_{\mathrm{s}}^2}{m_t^2}\,,\label{mh}\ee
where 
$m_{\tilde{t}_{1,2}}^2\sim m_q^2\sim m_u^2\gg m_t^2$ is assumed and 
additional terms coming from the stop mixing are neglected.\footnote{This expression
  is derived for the $\mathbb{Z}_3$ version of the NMSSM but the
  result holds for general versions of the NMSSM after replacing $\lambda \langle
  s\rangle $ for $\mu$.}
%$X_t=(A_t+\mu\cot\beta)$ is the mixing parameter in the stop-sector.
Notice that the tree level contribution (i.e.~the first two terms)
can be larger than in the MSSM for large $\lambda$ and low
$\tan\beta$, while the one loop correction (third term) depends
logarithmically on the SUSY scale $M_{\mathrm{s}}$ and is identical to the MSSM.

For completeness we provide the tree level  masses of the components of
the singlet, i.e.\ the 
scalar $m_{h_s}$, the pseudoscalar $m_{a_s}$ and
the fermion $m_{\chi_s}$. The expressions given below assume that there is no mixing in the mass matrices of the corresponding fields and were calculated following \cite{Ellwanger:2009dp} assuming non-vanishing $\mu_h$, see also \cite{Ross:2011xv}. They are given as follows
\be \begin{aligned}
m_{h_s}^2&=\frac{1}{2} \lambda A_{\lambda}\sin(2\beta)\frac{v^2}{s}+\kappa
s(A_{\kappa}+4\kappa s)-\lambda\mu_h\frac{v^2}{s}-\frac{\xi_s}{s}\ ,\\
m_{a_s}^2&=\frac{1}{2}\lambda(
A_{\lambda}+4\kappa s)\sin(2\beta)\frac{v^2}{s}-3 A_{\kappa}\kappa s-2b_s-\frac{\xi_s}{s}\ ,\\
m_{\chi_s}&=2\kappa s\ .
\end{aligned}\ee
   
\subsection{Calculation of $k$}\label{calculationk}

In this section we study  soft terms which naturally generate
the electroweak scale  within the
NMSSM. In particular,  we consider the following
non-uni\-versal  soft terms at the GUT scale
\be \begin{aligned}
&\quad m_0^2=m_{h_u}^2=m_{h_d}^2\ ,\quad
%\quad m_{q_3}^2=m_{u_3}^2=m_{d_3}^2=m_s^2=0,\\
\quad m_{q}^2=m_{u}^2=m_{d}^2=m_{l}^2=m_{e}^2=m_s^2=0,\\
&\quad M_0=M_{i=1,2,3}\ , \quad
\quad A_{u}=A_{\lambda}=A_{\kappa}=0\ ,\quad b_s=\xi_s=0\label{bc}\\
\end{aligned}\ee
while the parameters $b_{h_0}$ and $\mu_{h_0}$ are left free.\footnote{The index ``$0$''
denotes parameters which are taken at %the  GUT scale
$M_{\text{GUT}}\simeq
 10^{16}\text{GeV}$.}
Note that the parameters given in \eqref{bc} are flavor-diagonal  but
they are non-universal in that the 
soft Higgs masses differ from the soft sfermion masses.
%\footnote{ Here we differ from the scenario in \cite{Brummer:2013dya} where the third generation of quarks are also bulk fields.}.

To compute the soft terms at $M_{\rm s}$, the one-loop 
renormalization group equations (RGEs) are used \cite{Ellwanger:2009dp}
and  threshold corrections of the soft Higgs masses given in
\eqref{thresholdsofthiggs} are also included. The gauge couplings are
fixed at the GUT scale by
$\alpha_0=\alpha_2=\alpha_3=\frac{3}{5}\alpha_1\simeq0.04$ and 
only the top Yukawa $(y_t)$ and the  NMSSM Yukawa couplings 
$(\lambda$,$\kappa$) are taken into account while 
 all other Yukawa couplings are neglected. This approximation holds as long as $\tan\beta$ is not too large \cite{Martin:1997ns}.
 %Thus, the  Higgs scalar masses do not depend on any of the soft terms of the first and second generation. 
In sum, the free parameters before electroweak symmetry breaking are
\be M_0,\ m_0,\ \mu ,\  \tan\beta ,\  \lambda_0,\text{ and }
\kappa_0\ ,\label{parameters}\ee
where $\mu_h$ and $b_h$ have been traded for $\mu$ defined in \eqref{defmu} and $\tan\beta$ defined in \eqref{sin2b} respectively. 
From the RGE one obtains the soft Higgs mass parameters at low energy in terms of the GUT parameters. Explicitly one finds
%\be
%m_{h_{i=1,2}}^2= \alpha_i(y_{t0},\lambda_0,\kappa_0)\, M_0^2
%+\beta_i(y_{t0},\lambda_0,\kappa_0)\, m_0^2\label{RGEsofthiggs}\ ,
%\ee
\be
m_{h_{i=1,2}}^2= \alpha_i(\lambda_0,\kappa_0,\tan\beta)\, M_0^2
+\beta_i(\lambda_0,\kappa_0,\tan\beta)\, m_0^2\label{RGEsofthiggs}\ ,
\ee
where $\alpha_i,\beta_i$ are 
functions of the Yukawa couplings which can be computed numerically and we replaced the top Yukawa by $\tan\beta$ using \eqref{mtop}.

Using the assertion
\be M_0=k\, m_0\ ,\label{ansatz}\ee 
we computed the values of $k$  for which $\hat m$ in \eqref{m} is suppressed with respect to the supersymmetry breaking scale, i.e.
\be\hat m\ll M_{\mathrm{s}} \ .\ee
%Inserting the soft Higgs masses \eqref{RGEsofthiggs} and $m_0$ from \eqref{ansatz} into \eqref{m} one obtains\be \hat m^2={c}(y_{t0},\lambda_0,\kappa_0,\tan\beta,k)\ M_0^2\ ,\ee
Inserting the soft Higgs masses \eqref{RGEsofthiggs} and $m_0$ from \eqref{ansatz} into \eqref{m} one obtains\be \hat m^2={c}(\lambda_0,\kappa_0,\tan\beta,k)\ M_0^2\ ,\ee
where ${c}$ can be expressed in terms of $\alpha_i,\beta_i,k$ and $\tan\beta$. For the Yukawa coupling $\kappa$ at low energy we use $\kappa\sim0.4-0.6$ which corresponds to $\kappa_0\sim O(1)$.
% the result is almost not sensitive to the value of $\kappa$.
Thus, effectively $\hat m^2$ at low energy is parametrized by
\be \hat{m}^2=c(\lambda_0,\tan\beta,k)\ M_0^2.\ee

\begin{figure}[h!]
\centering\includegraphics[scale=0.6]{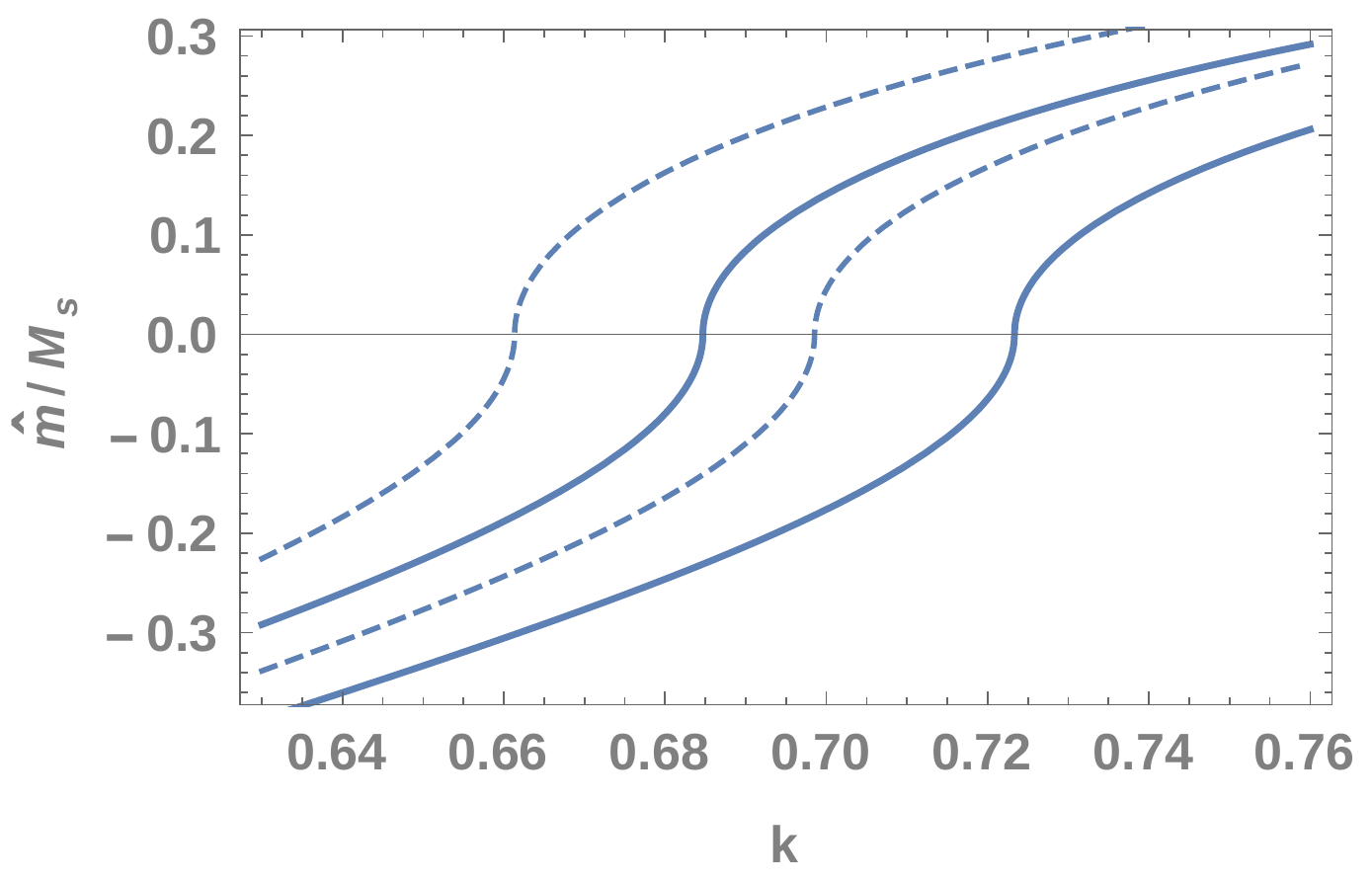}
\caption{$\hat{m}/M_s$ for different values of $\lambda_0$, from left
  to right $\lambda_0=0.4,0.001$, $\tan\beta=6,15$ (dashed,thick) and
  fixed $\kappa\simeq 0.4-0.6$, $M_{\mathrm{s}}=3\, \text{TeV}$. }
\label{fig:tanbvsalpha}
\end{figure}
%\begin{figure}[h!]
%\centering\includegraphics[scale=0.6]{tanbvsalpha.pdf}
%\caption{$\tan\beta$ vs $k$ with $\hat{m}\leq 400\text{GeV}$ for different values of $\lambda_0$ and fixed $\kappa\simeq 0.4-0.6$ and $M_{\mathrm{s}}=2\text{TeV}$. From left to right $\lambda_0=1.5,0.8,0.01$.}
%\label{fig:tanbvsalpha}
%\end{figure}

%In Figure \ref{fig:tanbvsalpha} we plot $\tan\beta$ versus $k$ for $\hat m\leq 400\text{GeV}$, $M_{\mathrm{s}}=2\text{TeV}$ and  $\lambda_0=1.5,0.8,0.01$ (from left to right).
% $\lambda_0\simeq1.5$ is the upper bound  to avoid Landau poles. 
 For $\lambda_0\ll1$ 
the singlet decouples and the Higgs sector is effectively the Higgs
sector of the MSSM. In this case the Higgs mass reaches its upper tree
level bound for large values of $\tan\beta$ and thus allows for
$M_{\mathrm{s}}={\cal O}(1 {\rm TeV})$.  From Figure
\ref{fig:tanbvsalpha} we see that in the regime  $\lambda_0\ll1$ and
for $0\lesssim \hat{m}\lesssim0.2\,M_s$, the range of the required $k$ 
%is independent of $\tan\beta$ and has to 
take values
in the narrow range
\be 0.70\lesssim k\lesssim 0.76\quad\text{(effective MSSM)}\ .\ee 

On the other hand, a phenomenologically interesting region in the
NMSSM corresponds to low $\tan\beta$ and large $\lambda_0$. In this
regime the tree level value of the Higgs mass is maximized and can
take larger values than in the MSSM case. However, too small values of
$\tan\beta$ imply a large cancellation of the two terms that contribute
to $\hat{m}$ in \eqref{m}, due to the fact that $m_{h_d}^2$ is large
at low energies. Hence we only consider moderate values of
$\tan\beta\, (\simeq 10)$, for which $\hat{m}^2\simeq -m_{h_u}^2$. Analogously, 
too large values of $\lambda_0$ induce  large $\mu_{\text{eff}}$, e.g. for $M_s=3\text{TeV}$ the upper bound $\lambda_0\lesssim 0.4$ corresponds to $\lambda\lesssim 0.33$ and $\mu_{\text{eff}}\lesssim500\text{GeV}$. Moreover, the upper bound on $\lambda_0$ is lessened for larger $M_s$.\footnote{The parameter $A_{\kappa}$ is negligible at low energy and thus can be disregarded in the calculation of $\langle s\rangle$. 
However, $\xi_s$ can get sizable radiative corrections provided  $\lambda_0$ is not too small. Similarly, $m_s^2$ and $b_s$ are the dominant contribution to $\langle s\rangle$ when $\lambda_0\to0$.
$\langle s\rangle$ is computed from \eqref{svev}.}
Notice that these constraints exclude the appealing regime of the NMSSM where the Higgs mass can get a larger tree level contribution.
Requiring $0\lesssim \hat{m}\lesssim0.2\,M_s$, $5\lesssim\tan\beta\lesssim15$ and
$0.01\lesssim\lambda_0\lesssim 0.4$ the range of $k$ widens 
\be 0.66\lesssim k\lesssim 0.76 \quad\text{(NMSSM)}.\label{rangek}\ee 

\begin{figure}%[htp!]
     \centering\includegraphics[scale=0.65]{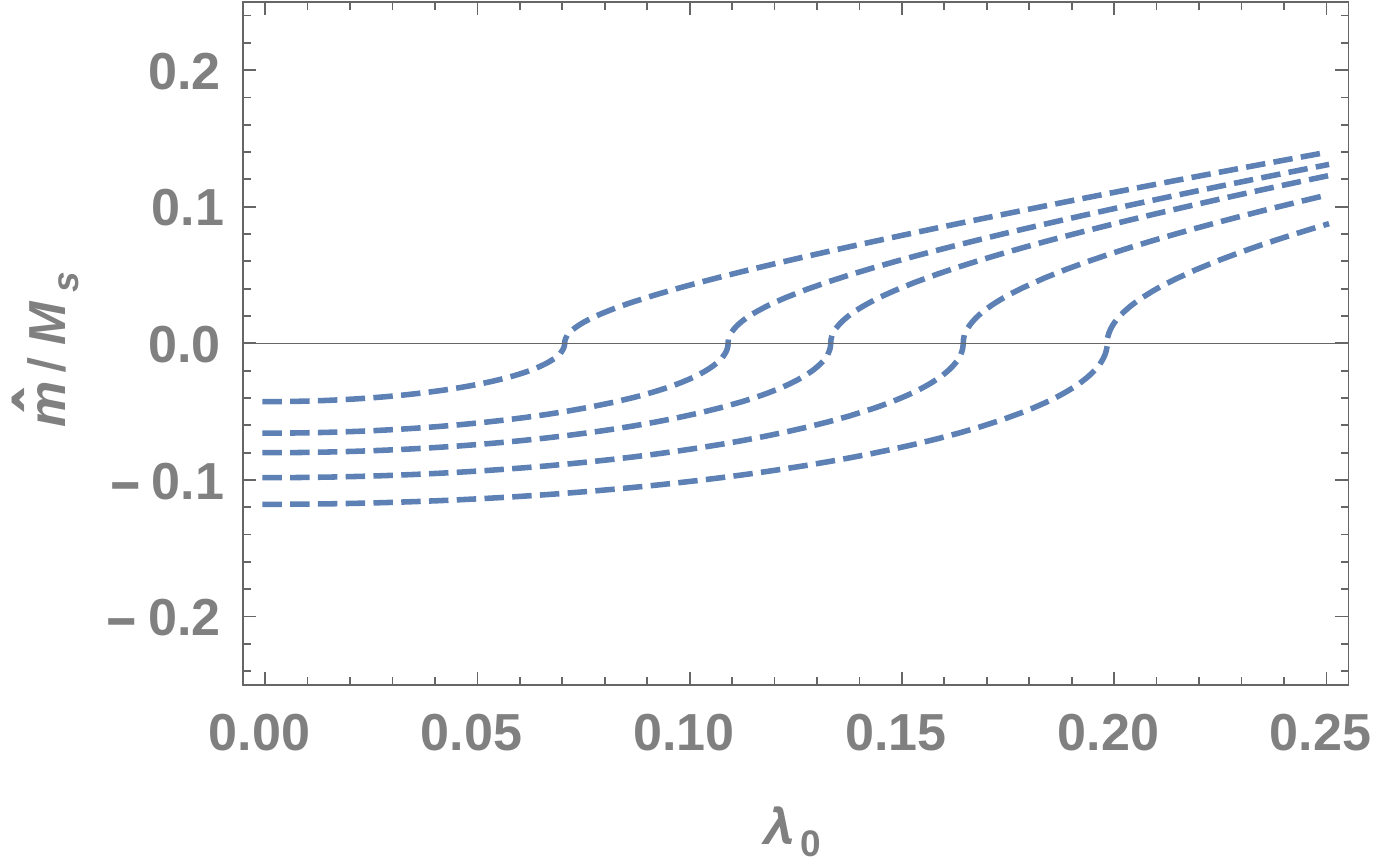}\\
     \vspace{1cm}
     \centering\includegraphics[scale=0.65]{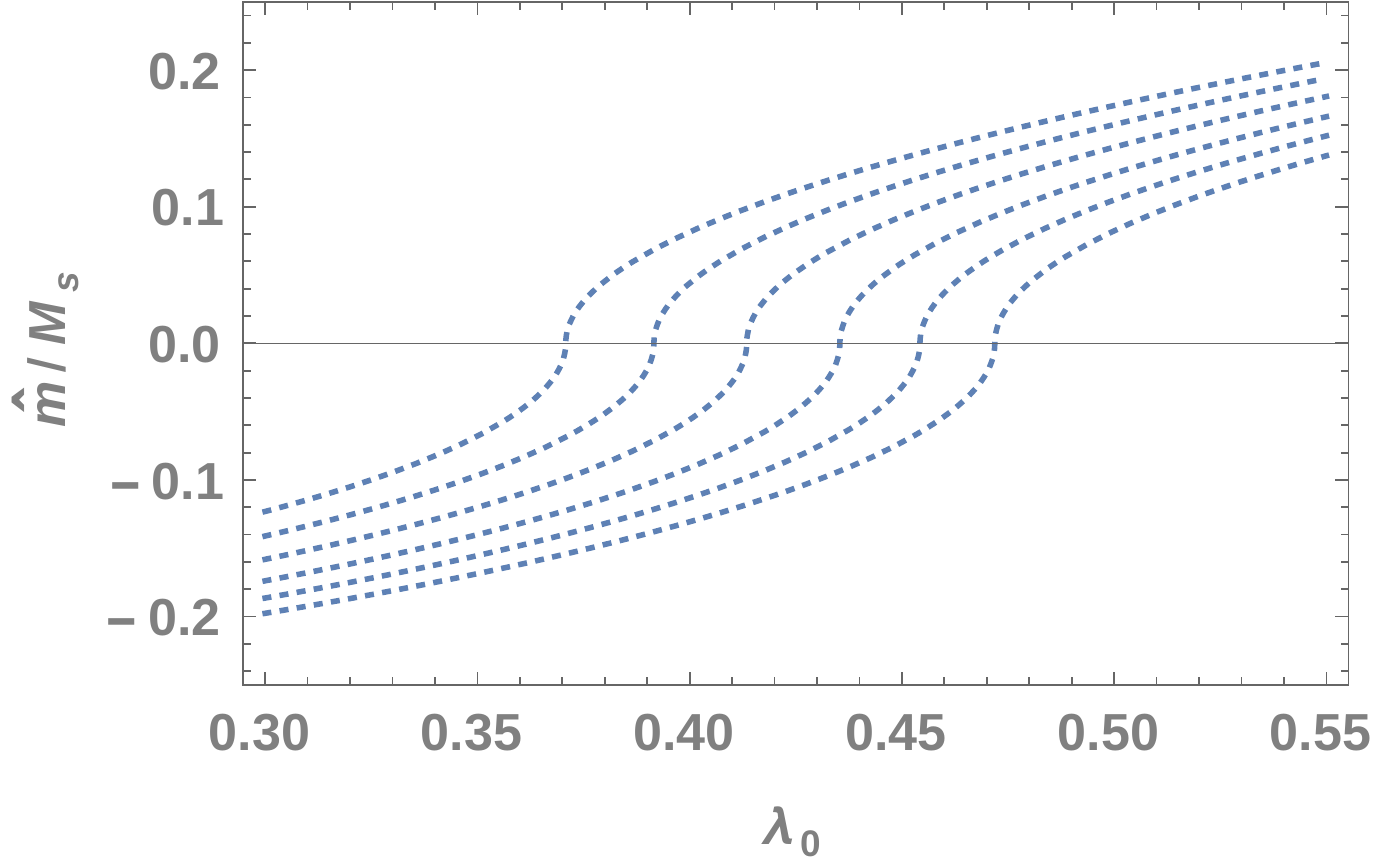}\\
\caption{ $\hat m/M_{\mathrm{s}}$  plotted as a function of $\lambda_0$, with $k=0.71$ (upper plot), $k=0.67$ (lower plot), increasing $\tan\beta$ between $6\text{ and }12$ from left to right and with  $M_{\mathrm{s}}\simeq 3\text{TeV}$. We see that  $\hat m$ is $\mathcal{O}(M_z)$ (region between -0.2 and 0.2) for a broad range of $\tan\beta$ and $\lambda_0$. }
\label{fig:mvsLam}
\end{figure}

%Notice from \eqref{defmu} that $\mu$ is a sum of two contributions and
%the value of $\mu_h$ can always be adjusted to yield
%$M_z\simeq91\text{GeV}$ in \eqref{Mz}. As we just discussed the relation
%\eqref{ansatz} 
%ensures $\hat m\ll M_{\mathrm{s}}$ for \eqref{rangek}, hence $\mu$ takes similar values. However, in order to 
%not introduce any additional fine-tuning between $\mu_\text{eff}$
%and $\mu_h$ we require $(\mu_{\text{eff}}\lesssim 450\text{GeV}$.
%This puts an upper bound on $\lambda_0$ of
% \be \lambda_0\lesssim0.8\,, \ee 
% i.e.\ $\lambda\simeq0.4$ at low scale, which does not depend on $k$.
% This value can be  decreased by taking lower values of $\kappa$. 
 %The upper bound can reach $\lambda_0\sim 1$ by taking larger values of $\kappa\sim0.6-0.7$ ($\kappa_0\sim3-4$). 
%
The $b_h$ parameter can be adjusted to give the desired values of
$\tan\beta$. Using \eqref{sin2b} and \eqref{rangek} the values of $b_h$ that give 
$5\lesssim\tan\beta\lesssim 20$ are within the range
% $b_h$ in gaugino mass units should lie within
\be 0\lesssim b_h/M_0^2\lesssim 0.4\,.\ee 
Finally, in Figure \ref{fig:mvsLam} we show that as promised
$\hat m\ll M_{\mathrm{s}}$ for  different values of
$\tan\beta$ and~$k$.

 %Let us finish this section by studying the constraint in the scale
 %invariant NMSSM, i.e.\ when $\mu_h=0$, and $b_h=0$. In this case,
 %$\mu$ and $\tan\beta$ in \eqref{parameters} are no longer free
 %parameters.  
 %In particular, $\mu$ is defined via \eqref{defmu} with
 %$\mu_h=0$ and it has to  obey $\mu\gtrsim100\,\text{GeV}$  to satisfy 
%the LEP bounds on charginos while $\tan\beta$ is determined by \eqref{sin2b} with $b_h=0$. 
%The calculation of $\hat{m}$ proceeds as before, i.e.\ we take the
%Yukawa coupling ($\kappa_0\simeq \mathcal{O}(1)$) at low energy $\kappa\sim0.4-0.6$ 
%and thus $\hat{m}$ is effectively parametrized by
%\be \hat{m}^2=c\,(\lambda_0,k)M_0^2\,.\ee 
%Electroweak symmetry breaking fixes $\lambda_0$ for each value of $k$, we take as before $M_s=2\text{TeV}$ and demand $100\,\text{GeV}\lesssim\mu\lesssim500\,\text{GeV}$. The range of $k$ in this case is in 
% the narrow window 
%\be 0.55\lesssim k\lesssim0.56 \quad\text{(scale invariant
%  NMSSM)},\label{rangek_nmssm}\ee 
%with $0.52\lesssim\lambda_0\lesssim 0.56$ and $\tan\beta\simeq20$. 
%{\bf Larger values of $\lambda_0$ give larger $\mu$ while lower values of $\lambda_0$ lead to larger values of $\tan\beta$ for which the bottom Yukawa coupling may become important.}

\subsection{Phenomenological implications}\label{pheno}

In this section we investigate the phenomenological implications of the scenarios studied in section \ref{calculationk}. In particular, depending on the value of $\lambda_0$, we find different predictions that could be tested in the next LHC run. 

Using \eqref{mh} we find that for
$M_{\mathrm{s}}\sim3-6 \text{TeV},\,
\tan\beta\simeq10$
the Higgs mass  is consistent with the measured value \cite{CMS:ril,ATLAS:2013sla} $m_h=125.6\text{GeV}$ within an uncertainty of $3\text{GeV}$
%\footnote{One can always take larger values of $M_s$ and hence of $M_0$ by going to lower values of $\tan\beta$ and/or $\lambda_0$. Thus the described spectrum should be understood as a lower bound on the susy particles.} 
 and we checked that the mixing of the singlet with
the Higgs is negligible in this range of parameters.
%\footnote{{\bf The mass matrix of the scalars was also computed in \cite{Ross:2011xv} for the GNMSSM. It can be applied for this setup after replacing $\mu_s=\xi=0$. The mixing between the Higgs and the singlet is proportional to $\lambda$ hence it becomes small for small values of $\lambda$. For larger values of $\lambda$ the diagonal entry of the singlet increases faster than the non-diagonal ones, hence the mixing is kept small.}}   

The above values of $M_{\mathrm{s}}$ correspond to $M_0\sim
1.5-3.5 \text{TeV}$.
The gluino mass $M_3$ obtained from the RGEs and stop masses calculated from \eqref{stops} are
\be {M}_3\sim4-8\text{TeV} \ ,\qquad
m_{\tilde{t}_{1,2}}\sim3-6\text{TeV}\ . \ee
From $M_3$ the wino and bino masses are computed via the standard 
relations \cite{Ellwanger:2009dp,Martin:1997ns}  $M_3:M_2:M_1\sim5.5,1.9,1$ giving
\be M_2\sim1400-3000  \,\text{GeV},\qquad M_1\sim700-1600 \,\text{GeV}\,.\ee

As we already discussed the 
Higgsino masses scale with $\mu$, which is bounded from below by $100\,\text{GeV}$. Since $\hat m$ is of the order of the electroweak scale we need to have $\mu$ in a similar range to obtain the correct Z boson mass.
As a consequence  the Higgsino masses turn out to be a few hundred GeV.
%The singlino is the heaviest neutralino with mass of $\mathcal{O}(\text{TeV})$.

\begin{figure}[h!]
  \begin{minipage}[b]{.51\linewidth}
\includegraphics[scale=0.54]{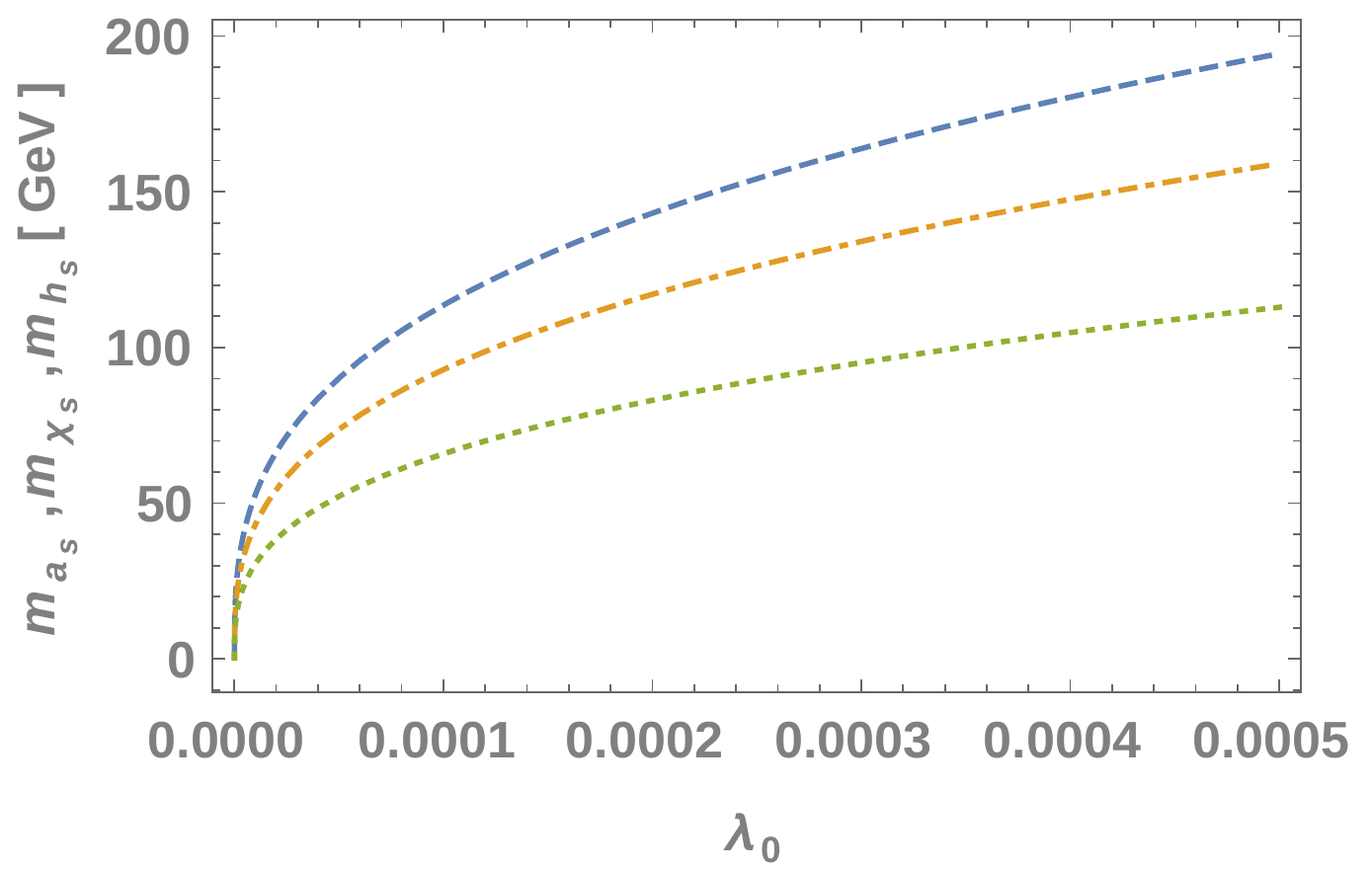}
  \end{minipage}%
  \begin{minipage}[b]{.51\linewidth}
\includegraphics[scale=0.54]{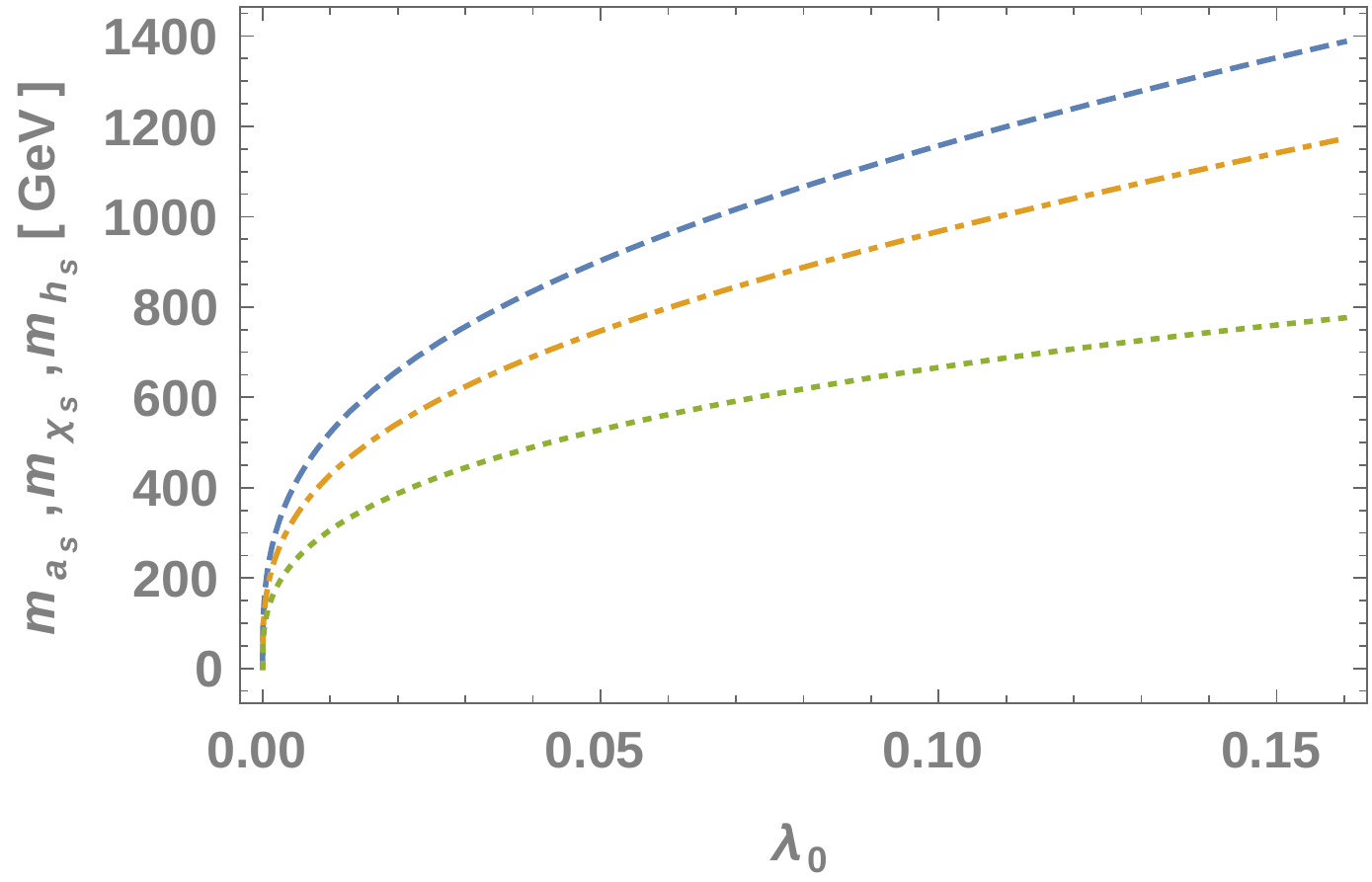}
  \end{minipage}
  \caption{
In the left figure we show the masses of the singlet scalar (dashed), singlino (dashed-dotted) and singlet pseudoscalar (dotted) for low values of $\lambda_0$ and $M_s=3\text{TeV}$. In the right plot the masses of the singlet and singlino are plotted for the same values of $M_\mathrm{s}$ and larger values of $\lambda_0$. One can see that they rapidly increase with $\lambda_0$. }
\label{fig:ms_lam}
\end{figure}

On the other hand, in the effective MSSM region we find that
 for very small $\lambda_0$ $\mathcal{O}(10^{-4})$ the neutral singlet can become lighter than the Higgs. In this regime, the singlino is the lightest neutralino (see Figure \ref{fig:ms_lam}). Moreover, a light singlet can yield significant changes in the Higgs decay constants that are consistent with the present LHC bounds. Experimental signatures have been recently studied and provide predictions for the next run \cite{Cao:2013gba,Curtin:2014pda,Curtin:2013fra}. For larger values
of $\lambda_0$ the singlet and singlino become heavy~$\mathcal{O}(\text{TeV})$.

The singlet pseudoscalar turns out to be also very light and its mass
strongly depends on the $\lambda_0$ coupling (see Figure \ref{fig:ms_lam}). In the scale invariant
version of the NMSSM, the potential
exhibits an approximate global $U(1)$ R-symmetry. This symmetry was
first discussed in \cite{Dobrescu:2000yn} and  it is exact at the GUT scale with $A_{\lambda}=A_{\kappa}=0$ as set in \eqref{bc} and becomes approximate at low energies via the radiative corrections to the  A-terms.
The symmetry is spontaneously broken when the scalars, $h_u,h_d\text{
  and }s$ get a VEV. 
The corresponding pseudo-Goldstone boson is the singlet pseudoscalar. 
Here the symmetry is already broken by the $\mu_h$ and $b_h$ terms terms at the GUT scale, however, provided that these together with $\lambda_0$ are small, the mass is slightly corrected from the scale invariant NMSSM case.
 Moreover, it can  modify the Higgs boson decays, the collider signatures of this scenario have been studied and are consistent with present LHC bounds \cite{Dermisek:2006wr,Cao:2013gba,Curtin:2014pda,Curtin:2013fra,Bomark:2014gya}.
 
%  \begin{figure}[h!]
%\centering\includegraphics[scale=0.75]{mps_lam.pdf}
%\caption{ Mass of the singlet pseudoscalar as a function of
%  $\lambda_0$ for different $M_s$ scales ($M_{\mathrm{s}}[\text{TeV}]=1,1.5,2,2.5$
%from bottom to top). The pseudoscalar mass vanishes for $\lambda_0\simeq0.05$ and remains light in all parameter space.}\label{fig:mps_lam}
%\end{figure}

 The spectrum of sleptons and squarks of the first and second generation resembles that of the MSSM in gaugino mediated scenarios \cite{Chacko:1999mi, Buchmuller:2005ma}. In particular, squarks are heavier than sleptons. The lightest sleptons are the right-handed ones and their masses lie below the bino neutralino within $m_{\tilde{e}_R}\simeq600-1300\,\text{GeV}$.
%The lightest is the right-handed component of the stau and its mass lies below the bino neutralino within $m_{\tilde{\tau}_R}\simeq220-500\text{GeV}$.}
 
% The mass of first and second generation of fermions could be higher,
% or lower, than the third generation of squarks at low energies
% depending on the boundary condition at the GUT scale. In the next
% section  we discuss  a higher dimensional model with the setup
% described in section \ref{calculationk}, where the boundary
% conditions have a geometrical interpretation. In section
% \ref{HDOG} %\ref{softterms} 
% we will see that the localisation of first and second
% generation of fermions could give rise to both possibilities  as pointed out in \cite{Brummer:2013dya}.

 We cross check the results with a modified version of SPheno \cite{Porod:2003um, Porod:2011nf} created by SARAH \cite{Staub:2008uz, Staub:2009bi, Staub:2010jh}.\footnote{We thank Kai Schmidt-Hoberg and Florian Staub for helping with the program.}
This performs a complete one-loop calculation of all SUSY and Higgs masses and includes the dominant two-loop corrections for the scalar Higgs masses. We show several benchmark points in Table \ref{tab:benchmark}, in particular, the spectrum for large $\lambda_0$ in P1 and P3, for small $\lambda_0$ in P2 and an intermediate value of $\lambda_0$ in $P4$.

\begin{table*}[h!t]
\centering
 \begin{tabular}{|l|c c c c|}
\hline

Parameters & P1 & P2 &  P3 &P4\\
\hline
\hline
$\lambda_0$             & 0.33            &     $10^{-4}$          &0.1  &$10^{-3}$\\
 $M_{0}$ [GeV]			   &  2000			&    2500        &3000    &3500\\
 $m_{0}^2$ [$\text{GeV}^2$]&  $7\cdot10^6$  &   $9.5\cdot10^6$  & $1.35\cdot10^{7}$  & $1.75\cdot10^{7}$\\
%$\tan \beta$            & 15              &        15       &  15& 15\\ 
%$\kappa_0$                            & 1           & 1   &   1  &  1 \\
\hline
%$\vert m_{h_u}\vert$ [GeV]             & 90              &               &     \\
%$\mu_h$  [GeV]                         &-90              &                 \\
%$b_h~[\text{GeV}^2]$       & $2.2\cdot10^5$  &       \\
%$\mu_{\text{eff}}$[GeV]             &  575            &                 \\
%$b_s~[\text{GeV}^2]$      & $-1.55\cdot10^5$        &     \\
%$\xi_s~[\text{GeV}^3]$   & $1.12\cdot 10^9$ &     \\
\hline  
$m_{h_s}$ [GeV]					&1850             		& 114.5  &    907.4      &  178.3 \\   
$m_{h}$ [GeV]					&123.6                 & 126   &    125.7      & 127.9\\   
$m_{H},m_{H^{\pm},m_{A}}$ [GeV]					&2824    &3434     &  4067     &    4660     \\   
%$m_{A_1}$ [GeV]                &2824   				&   &            \\  
$m_{a_s}$ [GeV]                & 1040       &      66.65    &  561    & 108.8 \\    
\hline
\hline
$m_{\tilde{\chi}_s}$ [GeV]          &  1659             & 93.65    &814.4 &147.8\\    
$m_{\tilde{\chi}_{\mu_1}}$ [GeV]           &491      & 695   &693	&766.2\\ 
$m_{\tilde{\chi}_{\mu_2}}$ [GeV]           &497      & 700   &696	&770\\ 
$m_{\tilde{\chi}_{\text{bino}}}$ [GeV]     &  880             &1106    &1335&1569\\     
$m_{\tilde{\chi}_{\text{wino}}}$ [GeV]      &   1642            &  2056  &2473 &2893\\     
\hline
\hline
$m_{\tilde{g}}$  [GeV]                 &  4070              & 5145       & 6104   & 7047\\      
$m_\text{squark}$ [GeV]					&  2680-3760         & 3330-4630      &3930-5480   & 4540-6310  \\     
$m_\text{slepton}$ [GeV]				&   667-1300            & 840-1620   &  1000-1940 &1180-2250\\     
%$m_{\tilde{\chi}^\pm_1}$ [GeV]			&  2820                 &    &            \\  
\hline  
\hline                         
 \end{tabular}
\caption{Examples of mass spectra computed with SPheno \cite{Porod:2003um, Porod:2011nf} created by SARAH \cite{Staub:2008uz, Staub:2009bi, Staub:2010jh}. We used $\tan\beta=15$, $\kappa_0=1$ and  $M_s=3,3.8,4.5,5\text{TeV}$ (from left ro right in the table).} 
\label{tab:benchmark}
\end{table*}

\section{Higher dimensional orbifold GUTs}\label{HDOG}

In this section we discuss an orbifold GUT  as an example which leads
to the previously chosen structure of soft terms in \eqref{bc} and the relation
\eqref{ansatz}. Let us start by discussing the low energy effective theory.

%\subsection{The setup}\label{setup}
 
The starting point is the assumption that the NMSSM is embedded in a
higher-dimen\-sional orbifold GUT. 
In these models a unified gauge group (e.g.\
$\text{SU}(5)\,\text{or}\,\text{SO}(10)$) in a higher-dimensional
theory
  breaks to the SM gauge group and $N=1$ supersymmetry in four
  space-time dimensions via the compactification
  \cite{Kawamura:2000ev,
  Altarelli:2001qj, Hebecker:2001wq, Asaka:2001eh, Hall:2001xr}. Thus, $M_{\text{GUT}}\simeq
  R^{-1}$ where $R$ is the compactification radius.
The four-dimensional low energy theory can be such that only the  MSSM (or extensions
thereof)  survive at scales below $R^{-1}$.
 An important aspect of these higher dimensional theories is that they
 are non-renormalizable, i.e perturbation theory can only be trusted
 up to a cutoff scale
$\Lambda$. 

In order to study phenomenological aspects, it
is necessary to specify where the matter content is localized and how
supersymmetry breaking occurs. We follow
ref.~\cite{Chacko:1999mi} (see also
  \cite{Chacko:1999hg,Kaplan:1999ac})
in that the gauge fields and MSSM Higgses live in the bulk whereas the
singlet and the $3^{\text{rd}}$ family of fermions sit at one of the
orbifold fix points. The $1^{\text{st}}$ and $2^{\text{nd}}$
fermion generations can be located at the same or a different fixed
point. %  which we do not specify for the moment.

We consider the situation where a hidden field $\Sigma$ sits at a fixed point
which is different from the fixed point of the singlet and the
$3^{\text{rd}}$ generation and
further assume that $\Sigma$  gets a VEV that triggers supersymmetry breaking \cite{Randall:1998uk}. 
$\Sigma$  is coupled through local universal operators to fields in
the bulk
which induce  soft breaking terms for the latter. However, soft terms for the fields that live on different, separated branes are suppressed at tree level by the size of the extra dimensions and can only develop radiatively.
For the localization of fields specified above it implies that soft
terms for the singlet and sfermions %third generation of squarks 
are negligible
while gaugino and Higgs soft scalar masses are sizable and
universal. With this setup the theory generates the soft terms in
\eqref{bc} and defines the boundary condition at the GUT scale.  Let
us now explicitly calculate $m_0$ and $M_0$ using the expressions given in Appendix \ref{appendixsoftterms}.\footnote{ One can consider the possibility that the $1^{\text{st}}$ and $2^{\text{nd}}$ generation of fermions sit at the same fixed point as the supersymmetry breaking field. In this case they could couple to the latter and get tree level soft terms. In particular, the soft scalar masses could be considerably heavier than the third generation of sfermions at low energy.}

%\subsection{Soft terms. Derivation of k}\label{softterms}
 
The soft terms arise from (non-renormalizable) couplings of the higher-dimensional theory
involving $\Sigma$, its $F$-term $F_\Sigma$ and the observable fields in the bulk. These couplings rely on the specific mechanism of supersymmetry breaking and they can be computed with the knowledge of the supersymmetric theory at high energies. At low energies the hidden sector decouples, thus the Lagrangian for observable fields reduces to the global supersymmetric piece plus soft terms. 
The leading order contribution to the soft terms is proportional  to the scale
\be m_{\text{soft}}=\frac{F_{\Sigma}}{\Lambda}\label{msoft}\ ,\ee
with $F_{\Sigma}$ being the supersymmetry breaking parameter. 
We follow the approach of \cite{Kaplunovsky:1993rd}, i.e.~do not specify the dynamics that triggers supersymmetry breaking and parametrize our ignorance through unknown couplings (functions of the hidden field) in the effective Lagrangian.
%This description relies on the fact that at low energies the principles of effective field theory can be applied.
Hence soft terms are  given in a model independent way  in terms of
the K\"{a}hler potential $(K)$, the superpotential $(W)$ and the gauge kinetic function $(f)$. 

Assuming that the couplings between the hidden field and the
observable fields in the bulk are universal we can parametrize
$K$, $W$ and $f$ as follows
\be\begin{aligned}
K&=Z(\hat{\Sigma},\hat{\bar{\Sigma}})\ (\,\vert H_u\vert^2+\vert
H_d\vert^2)+(\frac{1}{2}\mu_K\,(\hat{\Sigma},\hat{\bar{\Sigma}})H_u
H_d+\text{c.c.})\ ,\\%\mu_{k_2}\,(\hat{\bar{\Sigma}}^2H_uH_d+\text{c.c.})\ ,\\ 
f&=h(\hat{\Sigma})\ ,\quad\text{W}=\mu_{\text{W}}\,(\hat{\Sigma})H_u H_d\ ,
\end{aligned}\label{susyfc}\ee
where we defined
$\hat{\Sigma}=\frac{\Sigma}{\Lambda}$. 
Since we do not explicitly know the dynamics of the higher-dimensional
theory, the functions 
$Z,h,\mu_{K},\mu_{\text{W}}$ are unknown. 

Via the Giudice-Masiero mechanism \cite{Giudice:1988yz}
the last term in $K$ gives rise to the $\mu_h$ and $b_h$ parameters.
$\mu_h$ receives an additional contribution from
$\mu_{\text{W}}$ and thus does not have to be of order $
m_{\text{soft}}$. Therefore we treat $\mu_h$ as a free parameter
(i.e.\ do not address the $\mu$-problem).\footnote{See \cite{Lee:2011dya} for examples of effective $\mu_h$ terms after supersymmetry breaking in singlet extensions of the MSSM.}
%On the other hand, in the scale invariant NMSSM, the underlying $\mathbb{Z}_3$ symmetry forbids $\mu_\text{W}\text{ and }\mu_K$ terms, i.e $\mu_h=0$ and $b_h=0$, and thus there is no $\mu$ problem. 

After these preliminaries, the calculation of the soft gaugino and soft scalar masses is straightforward by means of \eqref{Msoft} and \eqref{msoft} in the appendix. Written in terms of canonically normalized fields, they are given as follows
%\be  M_0=\partial_{\Sigma} \log \text{Re}(h)\,m_{\text{soft}}\,,\quad m_{0}^2= -\partial_{\Sigma}\bar{\partial}_{\bar{\Sigma}}\log Z\,m_{\text{soft}}^2\label{softmasses}\ ,. \ee
\be  M_0=F_{\Sigma}\,\partial_{\Sigma} \log \text{Re}(h)\,,\quad m_{0}^2= -\vert F_{\Sigma}\vert^2\partial_{\Sigma}\bar{\partial}_{\bar{\Sigma}}\log Z\,\label{softmasses}\ . \ee  
From \eqref{softmasses} we derive the relation between the soft gaugino and soft higgs masses to be
\be  M_0=k\,m_{0}\,,\quad k=\frac{\partial_{\Sigma} \log \text{Re} (h)}{( -\partial_{\Sigma}\bar{\partial}_{\bar{\Sigma}}\log Z)^{\frac{1}{2}}}\ .\label{softk}\ee
Furthermore, the leading order contribution of this relation is obtained by expanding $Z$ and $h$ in powers of $\hat{\Sigma},\hat{\bar{\Sigma}}$ 
\be Z\simeq 1+ \rho\,\vert\hat{\Sigma}\vert^2+\ldots\ ,\quad h\simeq 1+\gamma \hat{\Sigma}+\ldots\ ,\label{expansion}\ee
% and for consistency we also expand $\mu_K$ and $\mu_W$
% \be ..\ee
where the numerical coefficients $\rho,\gamma$ are unknown constants.\footnote{The quadratic terms $(\Sigma^2+\hat{\bar{\Sigma}}^2)\ (\vert H_u\vert^2+\vert H_d\vert^2)$ give subleading contributions to the soft masses.} 
%Terms like $(1+\hat{\Sigma})H_uH_d+c.c.$ in the K\"{a}hler potential are neglected as they can be removed by a K\"{a}hler transformation.
A linear term in Z, $\rho_1\, (\hat{\Sigma} +\hat{\bar{\Sigma}})$ can be absorbed in $\rho$ by a field redefinition of the form $H_{u,d}\to (1+\rho_1)H_{u,d}$ and $\rho\to \rho-\rho_1^2$. Inserting \eqref{expansion} into \eqref{softk} yields
\be  k=\frac{1}{2}\frac{\gamma}{(-\rho)^{\frac{1}{2}}}\ .\ee
 
One example where $k$ can be estimated is using Na\"{i}ve Dimensional Analysis (NDA) \cite{Cohen:1997rt,Luty:1997fk}. NDA assumes that at the cutoff scale ($\Lambda$) all couplings, and their loop corrections, become order one in units of $\Lambda$, i.e.\ the theory becomes strongly coupled at energies near $\Lambda$. The corresponding ratio between gaugino and soft scalar masses in this case was computed in \cite{Brummer:2013dya} and is completely determined in a d-dimensional theory in terms of $\Lambda$ and the volume of the extra dimensions $V_{d-4}$. This relation is explicitly given by
\be M_0=k\,m_{0}\,,\quad \mathrm{with}\ k=\left(\frac{l_d}{l_4\Lambda^{d-4}V_{d-4}}\right)^{\frac{1}{2}},\label{softgauginos}\ee
with $l_d$ a numerical factor $l_d=2^d\pi^{d/2}\Gamma(d/2)$. 
 $\Lambda$ is bounded from above by the Planck scale in d-dimensions, which is defined via
\be M_{\mathrm{P},d}=\left(\frac{M_{\mathrm{P}}^2}{V_{d-4}}\right)^{\frac{1}{d-2}}. \ee
We calculate $k$ replacing $V=(2\pi R_l)^d$ with $R_l^{-1}\simeq M_{\text{GUT}}$ and $1.24 M_{\text{GUT}}\lesssim\Lambda\lesssim M_{\mathrm{P},d}$ (the lower bound of $\Lambda$ is determined by the absence of FCNC, see discussion below). For $d=5$ it yields 
\be 0.3\lesssim k\lesssim 0.8 \label{k5d}\ee
 and thus provides the coefficient in \eqref{rangek} with the expected size.  
For $d=6$ or larger the out coming $k$ is smaller than the required values. Notice that $k$ decreases with $\Lambda$, in particular, if $\Lambda$ takes the value of the Planck mass in 5 dimensions $k$ is too small to account for the necessary values.
 %Notice that if the cutoff is at the Planck scale, the value of $k$ (l.h.s in the bound) turns out too small. 
 %The Planck scale is the most natural choice for the cutoff, however, this option excludes the scenarios considered in this paper, i.e. the soft terms at the GUT scale defined by \eqref{bc} as in gaugino mediated schemes.  
%The MSSM models considered in \cite{Brummer:2013dya} do take the Planck mass as the cutoff scale but require the third generation of squarks to be in the bulk, at the expense of the solution to the flavour symmetry provided by gaugino mediated class of models. 
%Despite the fact that the window for $\Lambda$ is very constrained the obtained range of $k$ is larger than the range of $k$ computed in section \ref{calculationk}.  As already pointed out in the introduction the sensitivity of $M_z$ to $k$ is large, thus adjusting the value of $k$ implies a fine tuning as severe as in MSSM models. A precise determination of the  value of $k$ is crucial and to do so the complete UV theory is needed.}

Let us mention that FCNC in these models are absent as long as the cutoff is sufficiently large. More precisely, dangerous terms are generated through loops in the extra dimensions and scale like $\propto e^{-\Lambda L}$ with $L$ the distance between the branes \cite{Kaplan:1999ac}, here $L=2\pi R$. A suppression consistent with experimental bounds ($\lesssim 4\cdot 10^{-4}$) implies $\Lambda L\gtrsim7.8$, see \cite{Chacko:1999mi}. Thus, we must require a lower bound on $\Lambda$ of $\Lambda\gtrsim 1.24\,R^{-1}$. On the other hand, as stated above $\Lambda$ is bounded from above by $M_{P,d}$ which, for $V=(2\pi R_l)^d$ and $R_l^{-1}\simeq M_{\text{GUT}}$ in $d=5$, yields $\mathcal{O}(10^{17})\text{GeV}$ so the window for $\Lambda$ is quite constraint. 
%The range of $\Lambda$ could be relaxed in models with larger extra dimensions, i.e when $R$ is not tied to $M_{\text{GUT}}$.

Embedding the NMSSM into a spontaneously broken supergravity yields a
gravitino mass $m_{\frac{3}{2}}=\frac{F_{\Sigma}}{\sqrt{3}M_{\mathrm{P}}}$. The
relation between $\Lambda$ and $M_{\mathrm{P}}$ is model dependent,
however, as long as $\Lambda\ll M_{\mathrm{P}}$ the soft terms that
correspond to gravity mediated interactions are sub-leading and thus
can be neglected. Moreover, the gravitino mass generically appears as
the lightest supersymmetric particle (LSP) and is a good dark matter
candidate \cite{Buchmuller:2013dja}. For a study on gravitino dark
matter in gaugino mediation see \cite{Buchmuller:2006nx}. One can estimate $m_{\frac{3}{2}}$ by using $\Lambda$ as in the
calculation of soft terms for $d=5$ and $m_\text{soft}\simeq M_0$. This yields \be
m_{\frac{3}{2}}\simeq \mathcal{O}(0.006-0.06)\, M_0.\ee
Replacing $M_0$ as calculated in section~\ref{pheno} we find  $m_{\frac{3}{2}}\simeq 10-100\,\text{GeV}$  
and thus the gravitino can be the LSP. 
%depending on the value of $\lambda_0$ as studied in section \ref{pheno}.

\section{Conclusions}

In this paper we investigated a special relation among the soft terms that explains the small hierarchy between the supersymmetry breaking scale and the electroweak scale. More precisely, we looked for a condition of the soft terms for which the parameters that trigger electroweak symmetry breaking are suppressed with respect to the supersymmetry breaking scale. 
We considered the NMSSM and, as boundary conditions at the GUT scale, vanishing soft terms except of gaugino masses and soft terms for the Higgs sector. 
This setup can be embedded in e.g.~higher-dimen\-sional orbifold GUTs  where the singlet together with 
the quarks and leptons 
 are confined to a four dimensional subspace (brane or orbifold fixed
 point) while the  gauge fields and the Higgses propagate in the bulk.
Moreover, assuming that supersymmetry breaking occurs at a spatially separated brane by a hidden field leads to the so called gaugino mediation and yields the soft terms at the GUT scale studied in this paper.  
From the requirement of naturalness explained above we obtained a specific relation between the soft gauginos and Higgs scalar masses. 
 
%It is worth recalling that the regime in which the NMSSM has the appealing feature of enhancing the Higgs mass is not favored by this study, i.e.~small values of $\tan\beta$ and large values of $\lambda$ are excluded.  
% Moderate values of $\tan\beta$ are required to obtain a small $\hat{m}$ and too large values for $\lambda$ are not allowed since $\mu$ is required to be a small parameter.  
%Too small values of $\tan\beta$ and lareg values of $\lambda$ are excluded. 
%However, the window for the relation between soft gauginos and soft Higgs masses is widened with respect to the effective MSSM case when varying $\lambda$. 
In addition, we studied the phenomenology of these scenarios. The low energy spectrum depends on the value of the Yukawa coupling $\lambda$ of the singlet. For $\lambda\lesssim\mathcal{O}(10^{-1})$ the singlet scalar and singlino are  heavy $\mathcal{O}(\text{TeV})$ while for lower values $\lambda\lesssim \mathcal{O}(10^{-4})$ the singlet becomes the lightest scalar and the singlino the LSP. 
The pseudoscalar singlet is $\mathcal{O}(100) \text{GeV}$ below $\lambda\lesssim\mathcal{O}(10^{-1})$ and reaches $\mathcal{O}(\text{TeV})$ for larger values of $\lambda$.
%The pseudoscalar singlet is below $\mathcal{O}(100) \text{GeV}$ in all of the parameter space. 
The gravitino mass is $\mathcal{O}(10)\text{GeV}$ and can be the LSP depending on the value of $\lambda$. These scenarios can also be interesting for dark matter searches. 

Furthermore, we derived the soft terms for effective global
supersymmetric theories in a model independent way. This class of
theories can be considered as an intermediate step of the UV
completion and provide the necessary framework in e.g.~for gaugino
mediated scenarios. We specifically considered the example of
\cite{Brummer:2013dya} where higher-dimensional GUTs provide the special relation between soft gaugino and soft scalar masses required to explain the little hierarchy. This example relies on na\"{i}ve dimensional analysis arguments which allows to have explicit expressions for the couplings that determine the soft terms. In particular, these depend on the size of the extra dimensions and the cutoff scale and yield the required values for $d=5$ and a cutoff below the Planck mass.

\section*{Acknowledgements}
We would like to thank Felix Br\"{u}mmer,
David Ciupke, Florian Domingo, Fabian Ruehle, 
Georg Weiglein and especially  Willfried Buchm\"{u}ller and Kai
Schmidt-Hoberg for useful discussions. This work is supported by the German Science Foundation (DFG) within the Collaborative Research (CRC) 676  "Particles, Strings and the Early Universe".

\appendix
\section{Model independent soft terms in non renormalizable theories with global supersymmetry}\label{appendixsoftterms}

The starting point is a supersymmetric $N=1$ theory described by two
sectors: the {\it observable sector}, which includes (extensions of)
the  MSSM  and a  {\it hidden sector} that is responsible for
supersymmetry breaking. The chiral superfields in the observable
sector are denoted by $Q^{I}$ while the chiral fields in the hidden
sector are called $\phi^i$. Arbitrary non-renormalizable couplings are
allowed
which come suppressed by a cutoff scale $\Lambda$. 
%This theory should be understood as an effective field theory resulting from integrating out massive d.o.f that become relevant at the scale $\Lambda$.
The Lagrangian %of the supersymmetric theory 
can be completely specified in terms of the K\"{a}hler potential $K$,
the superpotential $W$ and the gauge kinetic function $f$. $K$ is a
real
and  gauge invariant and can be 
expanded in powers of the chiral fields $Q^{I},\bar{Q}^{\bar{I}}$
which reads
\be K=\Lambda^{2}
\hat{K}(\phi,\bar{\phi})+Z_{I\bar{J}}(\phi,\bar{\phi})Q^{I}\bar{Q}^{\bar{J}}+(\frac{1}{2}H_{IJ}(\phi,\bar{\phi})Q^IQ^J+c.c.)+\ldots\
.\ee
%where $\kappa:=1/\Lambda$ and the dots mean supressed higher order terms.
The superpotential is an holomorphic function of the chiral fields 
which is expanded as
\be
W(\phi,Q)=\hat{W}(\phi)+\frac{1}{2}\tilde{\mu}_{IJ}(\phi)Q^IQ^J+\frac{1}{3}Y_{IJK}(\phi)Q^IQ^JQ^K+...\
.\ee
%where in the trilinear terms it can be recognised the Yukawa
%interactions. 
The gauge kinetic function can depend on the hidden fields and defines the gauge couplings $g^{-2}_a(\phi,\bar{\phi})$ where  $a$ runs over different factors of the gauge group, i.e.\ 
$G=\prod_a G_a$. The $g_a$ renormalize in field theory with an all order expression given by \cite{Shifman:1986zi,Shifman:1991dz}
%\be g^{-2}_a(\phi,\bar{\phi})=\text{Re}f_a(\phi) + anomaly \ee
\be\begin{aligned}
g^{-2}_a(\phi,\bar{\phi},p)=\text{Re}f_a(\phi)&+\frac{b_a}{8\pi^2}\log\frac{\Lambda}{p}
+\frac{T(G_a)}{8\pi^2}\log
g^ {-2}_a(\phi,\bar{\phi},p)\\ &-\sum_{r}\frac{T_a(r)}{8\pi^2}\log\det
Z^{(r)}(\phi,\bar{\phi},p)\ .\end{aligned}\ee
Here $p<\Lambda$ is the renormalization scale and the numerical coefficients
are given by $T_a(r)=\text{Tr}_r(T_a^2)$, $
T(G_a)=T_a\text{(adjoint)}$
and $b_a = \sum_r n_r T_a(r) - 3 T(G_a)$ where 
 the summation is over representations $r$ of the (observable) gauge
 group G. The first term corresponds to the tree level gauge couplings
 while the other are loop corrections.

The effective potential of the hidden fields, responsible for supersymmetry breaking, can be written as
\be V^{\text{hid}}\simeq
\Lambda^{2}\hat{K}_{i\bar{j}}F^{i}\bar{F}^{\bar{j}}\ ,\ee
where 
\be \bar{F}^{\bar{j}}=\Lambda^{-2}\hat{K}^{\bar{j}i}\partial_i \hat{W}.\ee
Supersymmetry is spontaneously broken if $\langle F^{i}\rangle\neq0$ which defines the scale of SUSY breaking via
\be m_{\text{soft}}=\langle \hat{K}_{i\bar{j}}F^{i}\bar{F}^{\bar{j}}\rangle^{\frac{1}{2}}. \ee

Following \cite{Kaplunovsky:1993rd},  we calculate the effective
Lagrangian for the observable sector. In order to do so, we  replace
the hidden fields and their auxiliary partners by their VEVs. Keeping
only the renormalizable couplings we obtain
\begin{equation}\begin{aligned} V(Q,\bar{Q})=&\sum_a\frac{g_a^2}{4}(\bar{Q}^{\bar{I}}Z_{\bar{I}J}T_aQ^J)^2+\partial_I W^{\text{eff}}Z^{I\bar{J}}\bar{\partial}_{\bar{J}}\bar{W}^{\text{eff}}\\
&+m_{I\bar{J}}^2Q^{I}\bar{Q}^{\bar{J}}+(\frac{1}{3}A_{IJK}Q^IQ^JQ^K+\frac{1}{2}B_{IJ}Q^{I}Q^{J}+c.c.)\ ,\label{Veff}
\end{aligned}\end{equation}
where $W^{\text{eff}}$ denotes an effective superpotential defined as follows
\be
W^{\text{eff}}(Q)=\frac{1}{2}\mu_{IJ}Q^IQ^J+\frac{1}{3}Y_{IJK}Q^IQ^JQ^K\
,\ee
with \be \mu_{IJ}=\tilde{\mu}_{IJ}-\bar{F}^{\bar{j}}\bar{\partial}_{\bar{j}}H_{IJ}.\label{musoft}\ee
From \eqref{Veff} we see  that the first two terms correspond 
to a supersymmetric scalar potential while the last three terms are  
soft supersymmetry breaking terms. These soft terms depend on the original parameters of the K\"{a}hler function and the superpotential via
\begin{align}
m_{I\bar{J}}^2&=-F^{i}\bar{F}^{\bar{j}}R_{i\bar{j}I\bar{J}}\ ,\label{msoft}\\
A_{IJK}&=\ F^{i}D_iY_{IJK} \ ,\label{Asoft}\\
B_{IJ}&=\ F^{i}D_i\mu_{IJ}\ \label{Bsoft},
\end{align}
with
\begin{equation}\begin{aligned}
&R_{i\bar{j}I\bar{J}}=\partial_i\bar{\partial}_{\bar{j}}Z_{I\bar{J}}-\Gamma^N_{iI}Z_{N\bar{L}}\bar{\Gamma}^{L}_{\bar{j}\bar{J}}\,,\quad \Gamma^N_{iI}=Z^{N\bar{J}}\partial_iZ_{\bar{J}I}\ ,\\
&D_iY_{IJK} =\partial_iY_{IJK}-\Gamma^{N}_{i(I}Y_{JK)N}\ ,\\
&D_i\mu_{IJ} =\partial_i\mu_{IJ}-\Gamma^{N}_{i(I}\mu_{J)N}\ .
\end{aligned}\end{equation}
 From \eqref{msoft} one learns that this framework does not guarantee positive soft scalar masses, their sign is model dependent. Another observation is that, as in supergravity, $m_{I\bar{J}}$ need not be universal, hence the appearance of flavor mixing is also a problem in the global case. 

The kinetic term of the gauginos is given by
\be g_a^{-2}(\phi,\bar{\phi})\,\bar{\lambda}_a\sigma^{\mu} D_{\mu}\lambda_b\,.\ee
%and the soft gaugino mass term 
%\be M_a=-\frac{1}{2}F^n\partial_n f_a(\phi,\bar{\phi})\ .\ee
After canonically normalizing the kinetic term of gauginos the soft gaugino masses read
\be \frac{1}{2}(\,M_a\lambda^a\lambda^a+\ c.c.)\ ,\qquad M_a=F^i\partial_i\log g_a^{-2}(\phi,\bar{\phi}).\label{Msoft}\ee

\bibliographystyle{hieeetr}
\bibliography{Bibliography}

\begin{thebibliography}{10}

\bibitem{Aad:2012tfa}
G.~Aad {\em et~al.}, ``{Observation of a new particle in the search for the
  Standard Model Higgs boson with the ATLAS detector at the LHC},'' {\em
  Phys.Lett.}, vol.~B716, pp.~1--29, 2012, 1207.7214.

\bibitem{Chatrchyan:2012ufa}
S.~Chatrchyan {\em et~al.}, ``{Observation of a new boson at a mass of 125 GeV
  with the CMS experiment at the LHC},'' {\em Phys.Lett.}, vol.~B716,
  pp.~30--61, 2012, 1207.7235.

\bibitem{Chatrchyan:2013lya}
S.~Chatrchyan {\em et~al.}, ``{Search for supersymmetry in hadronic final
  states with missing transverse energy using the variables $\alpha_T$ and
  b-quark multiplicity in pp collisions at $\sqrt s=8$ TeV},'' {\em
  Eur.Phys.J.}, vol.~C73, no.~9, p.~2568, 2013, 1303.2985.

\bibitem{Aad:2013wta}
G.~Aad {\em et~al.}, ``{Search for new phenomena in final states with large jet
  multiplicities and missing transverse momentum at $\sqrt{s}$=8 TeV
  proton-proton collisions using the ATLAS experiment},'' {\em JHEP},
  vol.~1310, p.~130, 2013, 1308.1841.

\bibitem{Feng:2013pwa}
J.~L. Feng, ``{Naturalness and the Status of Supersymmetry},'' {\em
  Ann.Rev.Nucl.Part.Sci.}, vol.~63, pp.~351--382, 2013, 1302.6587.

\bibitem{Craig:2013cxa}
N.~Craig, ``{The State of Supersymmetry after Run I of the LHC},'' 2013,
  1309.0528.

\bibitem{Martin:1997ns}
S.~P. Martin, ``{A Supersymmetry primer},'' {\em Adv.Ser.Direct.High Energy
  Phys.}, vol.~21, pp.~1--153, 2010, hep-ph/9709356.

\bibitem{Horton:2009ed}
D.~Horton and G.~Ross, ``{Naturalness and Focus Points with Non-Universal
  Gaugino Masses},'' {\em Nucl.Phys.}, vol.~B830, pp.~221--247, 2010,
  0908.0857.

\bibitem{Feng:1999zg}
J.~L. Feng, K.~T. Matchev, and T.~Moroi, ``{Focus points and naturalness in
  supersymmetry},'' {\em Phys.Rev.}, vol.~D61, p.~075005, 2000, hep-ph/9909334.

\bibitem{Brummer:2013dya}
F.~Brummer and W.~Buchmuller, ``{A low Fermi scale from a simple gaugino-scalar
  mass relation},'' {\em JHEP}, vol.~1403, p.~075, 2014, 1311.1114.

\bibitem{Batra:2003nj}
P.~Batra, A.~Delgado, D.~E. Kaplan, and T.~M. Tait, ``{The Higgs mass bound in
  gauge extensions of the minimal supersymmetric standard model},'' {\em JHEP},
  vol.~0402, p.~043, 2004, hep-ph/0309149.

\bibitem{Ellwanger:2009dp}
U.~Ellwanger, C.~Hugonie, and A.~M. Teixeira, ``{The Next-to-Minimal
  Supersymmetric Standard Model},'' {\em Phys.Rept.}, vol.~496, pp.~1--77,
  2010, 0910.1785.

\bibitem{Lee:2011dya}
H.~M. Lee, S.~Raby, M.~Ratz, G.~G. Ross, R.~Schieren, {\em et~al.}, ``{Discrete
  R symmetries for the MSSM and its singlet extensions},'' {\em Nucl.Phys.},
  vol.~B850, pp.~1--30, 2011, 1102.3595.

\bibitem{Harigaya:2015iva}
K.~Harigaya, T.~T. Yanagida, and N.~Yokozaki, ``{Seminatural SUSY from $E_7$
  Nonlinear Sigma Model},'' 2015, 1504.02266.

\bibitem{Harigaya:2015jba}
K.~Harigaya, T.~T. Yanagida, and N.~Yokozaki, ``{Muon g-2 in Focus Point
  SUSY},'' 2015, 1505.01987.

\bibitem{Kawamura:2000ev}
Y.~Kawamura, ``{Triplet doublet splitting, proton stability and extra
  dimension},'' {\em Prog.Theor.Phys.}, vol.~105, pp.~999--1006, 2001,
  hep-ph/0012125.

\bibitem{Altarelli:2001qj}
G.~Altarelli and F.~Feruglio, ``{SU(5) grand unification in extra dimensions
  and proton decay},'' {\em Phys.Lett.}, vol.~B511, pp.~257--264, 2001,
  hep-ph/0102301.

\bibitem{Hebecker:2001wq}
A.~Hebecker and J.~March-Russell, ``{A Minimal S**1 / (Z(2) x Z-prime (2))
  orbifold GUT},'' {\em Nucl.Phys.}, vol.~B613, pp.~3--16, 2001,
  hep-ph/0106166.

\bibitem{Asaka:2001eh}
T.~Asaka, W.~Buchmuller, and L.~Covi, ``{Gauge unification in
  six-dimensions},'' {\em Phys.Lett.}, vol.~B523, pp.~199--204, 2001,
  hep-ph/0108021.

\bibitem{Hall:2001xr}
L.~J. Hall, Y.~Nomura, T.~Okui, and D.~Tucker-Smith, ``{SO(10) unified theories
  in six-dimensions},'' {\em Phys.Rev.}, vol.~D65, p.~035008, 2002,
  hep-ph/0108071.

\bibitem{Kobayashi:2004ud}
T.~Kobayashi, S.~Raby, and R.-J. Zhang, ``{Constructing 5-D orbifold grand
  unified theories from heterotic strings},'' {\em Phys.Lett.}, vol.~B593,
  pp.~262--270, 2004, hep-ph/0403065.

\bibitem{Forste:2004ie}
S.~Forste, H.~P. Nilles, P.~K. Vaudrevange, and A.~Wingerter, ``{Heterotic
  brane world},'' {\em Phys.Rev.}, vol.~D70, p.~106008, 2004, hep-th/0406208.

\bibitem{Kobayashi:2004ya}
T.~Kobayashi, S.~Raby, and R.-J. Zhang, ``{Searching for realistic 4d string
  models with a Pati-Salam symmetry: Orbifold grand unified theories from
  heterotic string compactification on a Z(6) orbifold},'' {\em Nucl.Phys.},
  vol.~B704, pp.~3--55, 2005, hep-ph/0409098.

\bibitem{Buchmuller:2007qf}
W.~Buchmuller, C.~Ludeling, and J.~Schmidt, ``{Local SU(5) Unification from the
  Heterotic String},'' {\em JHEP}, vol.~0709, p.~113, 2007, 0707.1651.

\bibitem{Randall:1998uk}
L.~Randall and R.~Sundrum, ``{Out of this world supersymmetry breaking},'' {\em
  Nucl.Phys.}, vol.~B557, pp.~79--118, 1999, hep-th/9810155.

\bibitem{Chacko:1999mi}
Z.~Chacko, M.~A. Luty, A.~E. Nelson, and E.~Ponton, ``{Gaugino mediated
  supersymmetry breaking},'' {\em JHEP}, vol.~0001, p.~003, 2000,
  hep-ph/9911323.

\bibitem{Chacko:1999hg}
Z.~Chacko, M.~A. Luty, and E.~Ponton, ``{Massive higher dimensional gauge
  fields as messengers of supersymmetry breaking},'' {\em JHEP}, vol.~0007,
  p.~036, 2000, hep-ph/9909248.

\bibitem{Schmaltz:2000ei}
M.~Schmaltz and W.~Skiba, ``{The Superpartner spectrum of gaugino mediation},''
  {\em Phys.Rev.}, vol.~D62, p.~095004, 2000, hep-ph/0004210.

\bibitem{Kaplunovsky:1993rd}
V.~S. Kaplunovsky and J.~Louis, ``{Model independent analysis of soft terms in
  effective supergravity and in string theory},'' {\em Phys.Lett.}, vol.~B306,
  pp.~269--275, 1993, hep-th/9303040.

\bibitem{Ellwanger:1993hn}
U.~Ellwanger, ``{Radiative corrections to the neutral Higgs spectrum in
  supersymmetry with a gauge singlet},'' {\em Phys.Lett.}, vol.~B303,
  pp.~271--276, 1993, hep-ph/9302224.

\bibitem{Ross:2011xv}
G.~G. Ross and K.~Schmidt-Hoberg, ``{The Fine-Tuning of the Generalised
  NMSSM},'' {\em Nucl.Phys.}, vol.~B862, pp.~710--719, 2012, 1108.1284.

\bibitem{CMS:ril}
``{Updated measurements of the Higgs boson at 125 GeV in the two photon decay
  channel},'' 2013.

\bibitem{ATLAS:2013sla}
``{Combined coupling measurements of the Higgs-like boson with the ATLAS
  detector using up to 25 fb$^{-1}$ of proton-proton collision data},'' 2013.

\bibitem{Cao:2013gba}
J.~Cao, F.~Ding, C.~Han, J.~M. Yang, and J.~Zhu, ``{A light Higgs scalar in the
  NMSSM confronted with the latest LHC Higgs data},'' {\em JHEP}, vol.~1311,
  p.~018, 2013, 1309.4939.

\bibitem{Curtin:2014pda}
D.~Curtin, R.~Essig, and Y.-M. Zhong, ``{Uncovering light scalars with exotic
  Higgs decays to bbmumu},'' 2014, 1412.4779.

\bibitem{Curtin:2013fra}
D.~Curtin, R.~Essig, S.~Gori, P.~Jaiswal, A.~Katz, {\em et~al.}, ``{Exotic
  decays of the 125 GeV Higgs boson},'' {\em Phys.Rev.}, vol.~D90, no.~7,
  p.~075004, 2014, 1312.4992.

\bibitem{Dobrescu:2000yn}
B.~A. Dobrescu and K.~T. Matchev, ``{Light axion within the next-to-minimal
  supersymmetric standard model},'' {\em JHEP}, vol.~0009, p.~031, 2000,
  hep-ph/0008192.

\bibitem{Dermisek:2006wr}
R.~Dermisek and J.~F. Gunion, ``{The NMSSM Close to the R-symmetry Limit and
  Naturalness in..},'' {\em Phys.Rev.}, vol.~D75, p.~075019, 2007,
  hep-ph/0611142.

\bibitem{Bomark:2014gya}
N.-E. Bomark, S.~Moretti, S.~Munir, and L.~Roszkowski, ``{A light NMSSM
  pseudoscalar Higgs boson at the LHC redux},'' 2014, 1409.8393.

\bibitem{Buchmuller:2005ma}
W.~Buchmuller, J.~Kersten, and K.~Schmidt-Hoberg, ``{Squarks and sleptons
  between branes and bulk},'' {\em JHEP}, vol.~0602, p.~069, 2006,
  hep-ph/0512152.

\bibitem{Porod:2003um}
W.~Porod, ``{SPheno, a program for calculating supersymmetric spectra, SUSY
  particle decays and SUSY particle production at e+ e- colliders},'' {\em
  Comput.Phys.Commun.}, vol.~153, pp.~275--315, 2003, hep-ph/0301101.

\bibitem{Porod:2011nf}
W.~Porod and F.~Staub, ``{SPheno 3.1: Extensions including flavour, CP-phases
  and models beyond the MSSM},'' {\em Comput.Phys.Commun.}, vol.~183,
  pp.~2458--2469, 2012, 1104.1573.

\bibitem{Staub:2008uz}
F.~Staub, ``{SARAH},'' 2008, 0806.0538.

\bibitem{Staub:2009bi}
F.~Staub, ``{From Superpotential to Model Files for FeynArts and
  CalcHep/CompHep},'' {\em Comput.Phys.Commun.}, vol.~181, pp.~1077--1086,
  2010, 0909.2863.

\bibitem{Staub:2010jh}
F.~Staub, ``{Automatic Calculation of supersymmetric Renormalization Group
  Equations and Self Energies},'' {\em Comput.Phys.Commun.}, vol.~182,
  pp.~808--833, 2011, 1002.0840.

\bibitem{Kaplan:1999ac}
D.~E. Kaplan, G.~D. Kribs, and M.~Schmaltz, ``{Supersymmetry breaking through
  transparent extra dimensions},'' {\em Phys.Rev.}, vol.~D62, p.~035010, 2000,
  hep-ph/9911293.

\bibitem{Giudice:1988yz}
G.~Giudice and A.~Masiero, ``{A Natural Solution to the mu Problem in
  Supergravity Theories},'' {\em Phys.Lett.}, vol.~B206, pp.~480--484, 1988.

\bibitem{Cohen:1997rt}
A.~G. Cohen, D.~B. Kaplan, and A.~E. Nelson, ``{Counting 4 pis in strongly
  coupled supersymmetry},'' {\em Phys.Lett.}, vol.~B412, pp.~301--308, 1997,
  hep-ph/9706275.

\bibitem{Luty:1997fk}
M.~A. Luty, ``{Naive dimensional analysis and supersymmetry},'' {\em
  Phys.Rev.}, vol.~D57, pp.~1531--1538, 1998, hep-ph/9706235.

\bibitem{Buchmuller:2013dja}
W.~Buchmuller, V.~Domcke, K.~Kamada, and K.~Schmitz, ``{A Minimal
  Supersymmetric Model of Particle Physics and the Early Universe},''
  pp.~47--77, 2013, 1309.7788.

\bibitem{Buchmuller:2006nx}
W.~Buchmuller, L.~Covi, J.~Kersten, and K.~Schmidt-Hoberg, ``{Dark Matter from
  Gaugino Mediation},'' {\em JCAP}, vol.~0611, p.~007, 2006, hep-ph/0609142.

\bibitem{Shifman:1986zi}
M.~A. Shifman and A.~Vainshtein, ``{Solution of the Anomaly Puzzle in SUSY
  Gauge Theories and the Wilson Operator Expansion},'' {\em Nucl.Phys.},
  vol.~B277, p.~456, 1986.

\bibitem{Shifman:1991dz}
M.~A. Shifman and A.~Vainshtein, ``{On holomorphic dependence and infrared
  effects in supersymmetric gauge theories},'' {\em Nucl.Phys.}, vol.~B359,
  pp.~571--580, 1991.

\end{thebibliography}

\end{document}